\documentclass{aa}  

\usepackage{graphicx}
\usepackage{txfonts}
\usepackage{lscape}
\usepackage{caption}
\usepackage{longtable}
\usepackage{xcolor}
\usepackage{rotating}
\usepackage{float}
\usepackage{subfig}
\usepackage{multicol}
\usepackage{multirow, array}
\usepackage{graphicx}
\usepackage{natbib}
\usepackage[titletoc,toc,page]{appendix}
\usepackage{amstext}

\begin{document}

   \title{Relation between metallicities and spectral energy distributions of Herbig Ae/Be stars. A potential link with planet formation}

   \subtitle{}
   
\titlerunning{Metallicities of Herbig Ae stars in relation with spectral energy distributions}
   
   \author{J. Guzmán-Díaz\inst{1}\and B. Montesinos\inst{1}\and I. Mendigutía\inst{1}\and M. Kama\inst{2}\and G. Meeus\inst{3}\and M. Vioque\inst{4,5}\and R.D. Oudmaijer\inst{6}\and E. Villaver\inst{1}
          }

   \institute{$^{1}$Centro de Astrobiología (CSIC-INTA), ESA-ESAC Campus, 28692, Villanueva de la Cañada, Madrid, Spain\\$^{2}$Department of Physics and Astronomy, University College London, Gower Street, London, WC1E 6BT, UK\\$^{3}$Departamento Física Teórica, Facultad de Ciencias, Universidad Autónoma de Madrid, Campus de Cantoblanco, 28049, Madrid, Spain\\$^{4}$Joint ALMA Observatory, Alonso de Córdova 3107, Vitacura, Santiago 763-0355, Chile\\$^{5}$National Radio Astronomy Observatory, 520 Edgemont Road, Charlottesville, VA 22903, USA\\$^{6}$ School of Physics and Astronomy, University of Leeds, Leeds LS2 9JT, UK
             }

   \date{Received 10 November 2022; accepted 22 December 2022 }

% \abstract{}{}{}{}{} 
% 5 {} token are mandatory
 
  \abstract
  % context heading (optional)
  % {} leave it empty if necessary  
   {Most studies devoted to Herbig Ae/Be stars (HAeBes) assume solar metallicity. However, the stellar metallicity, [M/H], is a fundamental parameter that can strongly differ depending on the source, and may have important implications for planet formation. In particular, Kama et al. proposed that the deficit of refractory elements observed in the surfaces of some HAeBes may be linked to the presence of cavities in their disks, and is likely caused by Jovian planets that trap the metal-rich content.}
  % aims heading (mandatory)
   {This work aims to provide a robust test on the previous proposal by analyzing the largest sample of HAeBes with homogeneously derived [M/H] values, stellar, and circumstellar properties.}
  % methods heading (mandatory)
   {Spectra of 67 HAeBes with well known properties from our previous work have been collected from the ESO Science Archive Facility. Their [M/H] values have been derived based on the comparison with Kurucz synthetic models. Statistical analyses have been carried out aiming to test the potential relation between [M/H] and the group I sources from the spectral energy distribution (SED) classification by Meeus et al., whose disks have been associated to the presence of cavities potentially carved by giant planets. We have critically analyzed the eventual link between [M/H], the SED groups, and the presence of such planets.}
  % results heading (mandatory)
   {Our statistical study robustly confirms that group I sources tend to have a lower [M/H] (typically $\sim -0.10$) than that of group II HAeBes ($\sim$ +0.14). A similar analysis involving SED-based transitional disks -with infrared excess only at wavelengths $\geq$ 2.2 $\mu$m- does not reveal such a relation with [M/H], indicating that not all processes capable of creating holes in the inner dust disks have an effect on the stellar abundances. The spatial distributions of group I and II sources are similar, at least within the available range of distances to the galactic centre and the galactic plane, for which the observed [M/H] differences are not driven by environmental effects. In addition, group I sources tend to have stronger (sub-) mm continuum emission presumably related to the presence of giant planets. Indeed, literature results indicate that disk substructures probably associated to the presence of giant planets are up to ten times more frequent in group I HAeBes than in group II. Finally, along with the metallicities derived for the whole sample, surface gravities and projected rotational velocities are additional outcomes of this work.}
  % conclusions heading (optional), leave it empty if necessary 
   {We provide indirect evidences suggesting that giant planets are more frequent around group I/low [M/H] stars than around the rest of the HAeBes. However, the direct test of the previous hypothesis requires multiple detections of forming planets in their disks. Such detections are so far limited to the candidate around the metal depleted ([M/H] = $-0.35\pm 0.25$) group I HAeBe star AB Aur, consistent with our findings.}


   \maketitle
%
%-------------------------------------------------------------------

\section{Introduction}\label{intro}

Stellar metallicity, [M/H]\footnote{[M/H] is defined as $\log\,[{\rm N}_{\rm M}/{\rm N}_{\rm H}]_{\rm star} - \log\,[{\rm N}_{\rm M}/{\rm N}_{\rm H}]_{\odot}$, ${\rm N}_{\rm M}$ and ${\rm N}_{\rm H}$ being the abundances of all elements heavier than hydrogen and helium, and the abundance of hydrogen, respectively.}, may play a fundamental role in planet formation. Soon after the discovery of the first exoplanets around solar-type stars, a trend linking the presence  of planets to stars with higher [M/H] values became clear (\citealt{Gonzalez_1997}; \citealt{Santos_2000,Santos_2001}). As the  number of detected exoplanets increased, further evidence confirmed and strengthen the today well-known "planet-metallicity" correlation for FGK stars (see e.g. \citealt{Adibekyan_2019}, \citealt{Osborn_2020}, and references therein). In turn, the situation for A-type stars is far from clear. Less than 20 planets have been confirmed around stars with effective temperatures between 8000 and 12000 K\footnote{http://exoplanet.eu/}, which, at this stage, makes a robust statistical study unfeasible. Indeed, the most common methods to detect exoplanets are based on weak spectroscopic velocity signals and photometric transits, for which the relatively small number of photospheric signatures, typically large rotational velocities and high luminosities of hot stars make them difficult targets to probe. 

Concerning the precursors of main-sequence A and B stars, recent studies have been carried out to characterize large samples of young, intermediate-mass Herbig Ae/Be objects \citep[HAeBes,][]{Vioque_2018,Arun_2019,Wichittanakom20,Guzman-Diaz_2021} and to significantly increase the number of potential members belonging to the class \citep{Vioque20,Vioque22,Zhang2022,Kuhn2022}. However, most studies devoted to HAeBes do not usually analyze [M/H], and the solar value is adopted by default. Only a few  works have considered the chemical peculiarities that HAeBes can show \citep[such as the $\lambda$ Bo{\"o}tis phenomenon, see e.g.][]{Gray_1998}, determining abundances for relatively small sub-samples (e.g. \citealt{AckeWaelkens_2004}, \citealt{Guimaraes_2006}, \citealt{Montesinos09}, \citealt{Folsom_2012}). 

A major study involving [M/H] in HAeBes was carried out by \cite{Kama15}. They proposed that the HAeBes showing a deficit of refractory elements, and thus low values of [M/H], could be linked to the presence of Jupiter-like size exoplanets in their protoplanetary disks. This scenario does not necessarily imply that A and B stars harboring giant planets are globally metal-poor, which would be opposite to the planet-metallicity correlation found in late-type stars. According to \cite{Kama15}, forming planets mainly trap the metal-rich material, the remaining accreted by the central star being metal-depleted. Given that A and B stars have radiative envelopes, the mixing timescale with the interior is of the order of $\sim$ Myr, much slower than that for lower mass stars with convective sub-photospheric regions. Therefore, [M/H] measurements in HAeBes refer to freshly accreted material and only reflect the stellar surfaces. Additional details on the theoretical model behind the previous scenario are described in \citet{Jermyn_2018}, and similar views proposed to explain the properties of some post-AGB stars and solar twins can be consulted e.g. in \citet{Oomen2019}, \citet{Booth2020}, \citet{Kluska_2022} and references therein. Other scenarios involving changes in the stellar metallicity have been explored e.g. in the context of stars with planets and debris disks \citep{Maldonado2012,Maldonado2015} or at galactic scales \citep[e.g.][and references therein]{Adibekyan2014,Hawkins_2022}.  

On the other hand, the proposal by \cite{Kama15} hinges on the fact that the deficit in [M/H] is observed in stars belonging to  group I in the classification by \citet{Meeus_2001}. This is based on the shape of the spectral energy distribution (SED) in the infrared and submillimeter regions. Such a classification has been related, among others, with the morphology  of the circumstellar disks, where group I and group II sources are associated with "flared" and self-shadowed, "flattened" disks, respectively (e.g. \citealt{Meeus_2001}, \citealt{Dullemond02}, \citealt{Dullemond04}). Most relevant to the hypothesis by \cite{Kama15}, studies based on high-resolution imaging suggest that disks belonging to group I sources show cavities potentially carved out by giant planets (e.g. \citealt{Maaskant13}, \citealt{Honda_2015}, \citealt{Garufi17}, \citealt{Stapper_2021}).

Although the scenario proposed by \citet{Kama15} could have major implications for our understanding of planet formation, that was based on a study of a relatively small sample of 22 HAeBes. In fact, these were the only ones having measured [M/H] values and disk structure classification from SEDs by that time, which were based on heterogeneous studies from the literature. Our main aim is to solve these issues by providing a robust test on whether the [M/H] of HAeBes can be related to a specific SED-based disk structure from the \citet{Meeus_2001} scheme or not. Given that the answer we find is affirmative, the potential implications in terms of the presence of giant planets are further analyzed. The departure point is our previous work in \citet{Guzman-Diaz_2021}, where the stellar parameters and the circumstellar morphology in terms of the \citet{Meeus_2001} groups were homogeneously derived for essentially all classical, historically well-know HAeBes. Here we derive [M/H] for all such objects where this parameter can be measured from suitable spectra, increasing the sample analyzed by \citet{Kama15} by a factor $\sim$ 3. On top of the new [M/H] values for all the stars, projected rotational velocities ($v \sin i$) and surface gravities ($\log g$) will also be provided as valuable outcomes of our work. Section \ref{sample} describes the sample and the observations. The process to estimate [M/H] and the rest of stellar parameters is described in Sect. \ref{metallicity}. An analysis of the previous results is in Sect. \ref{results}, which includes a statistical study and discussion regarding the connection between [M/H] and the SED classification. Finally, a brief summary and the main conclusions are in Sect. \ref{conclusions}.

%--------------------------------------------------------------------

\section{Sample and observations}\label{sample}

HAeBes with effective temperatures, T{\tiny eff}, below 12000 K (spectral types later than B8) from \citet{Guzman-Diaz_2021} and with one-dimensional spectra available at the ESO Science Archive Facility\footnote{http://archive.eso.org/cms.html} were selected. Accordingly, the stellar mass ranges typically between 1 and 5 M$_{\odot}$ and the ages are mostly $<$ 15 Myr. We discarded emission-dominated stars not showing enough absorption lines (i.e. mainly HBe stars with T{\tiny eff} $>$ 12000 K). Sources with spectra not having a large enough signal-to-noise ratio (SNR typically $\geq$ 100 for all stars except for four listed in Appendix \ref{appB}, with SNR $\geq$ 50) or spectral resolution (see below) were also discarded. The final sample is constituted by 67 sources, whose main properties are summarized in Table \ref{Table1}. All data listed in that table were compiled from the photometric study of \citet{Guzman-Diaz_2021}, except for the spectroscopically determined T{\tiny eff} values from \citet{Wichittanakom20} whenever available. The sample covers a range of luminosities -in log(L$_*$/L$_\odot$)- between 0.4 and 3.2, masses between 1.4 and 7.0 M$_{\odot}$, and ages between 0.2 and 20 Myr, representing almost 50\% of all classical, late type HAeBes studied in \citet{Guzman-Diaz_2021}. Regarding the SEDs studied in that work, 28 sources were classified as group I and 34 as group II. Although five stars could not be classified concerning the SED groups, they have also been included in the current sample for completeness, given that their [M/H] values could also be determined from the available spectra.   

Most spectra were taken with the XSHOOTER/VLT spectrograph. The spectral resolution, R = $\lambda/\Delta\lambda$, varies between $\sim\!3200$ and $\sim\!18\,400$ depending on the wavelength range and the slit width. Spectra taken with UVES/VLT and HARPS/La Silla 3.6-m, were also used, with R $\sim\!41\,000-110\,000$ and 
$\sim\!80\,000-115\,000$, respectively. Finally, the spectrum of HBC 222 was taken with GIRAFFE/VLT (R $\sim\!5500-65\,000$). The spectrographs used to obtain the spectra of each source, and their resolutions at $\sim$ 5000 \r{A} are listed in Cols. 2 and 3 of Table \ref{Table2}. The first initials indicate the instruments and resolutions corresponding to the spectra used to estimate $v \sin i$, and the second to those used for measuring the widths of the Balmer lines (i.e. $\log g$) and [M/H] (see Sect. \ref{metallicity}). In case several spectra from the same instrument are available for a given star, priority was given to the one with the highest spectral resolution.

\begin{figure*}
 \centering
    \includegraphics[width=0.80\textwidth]{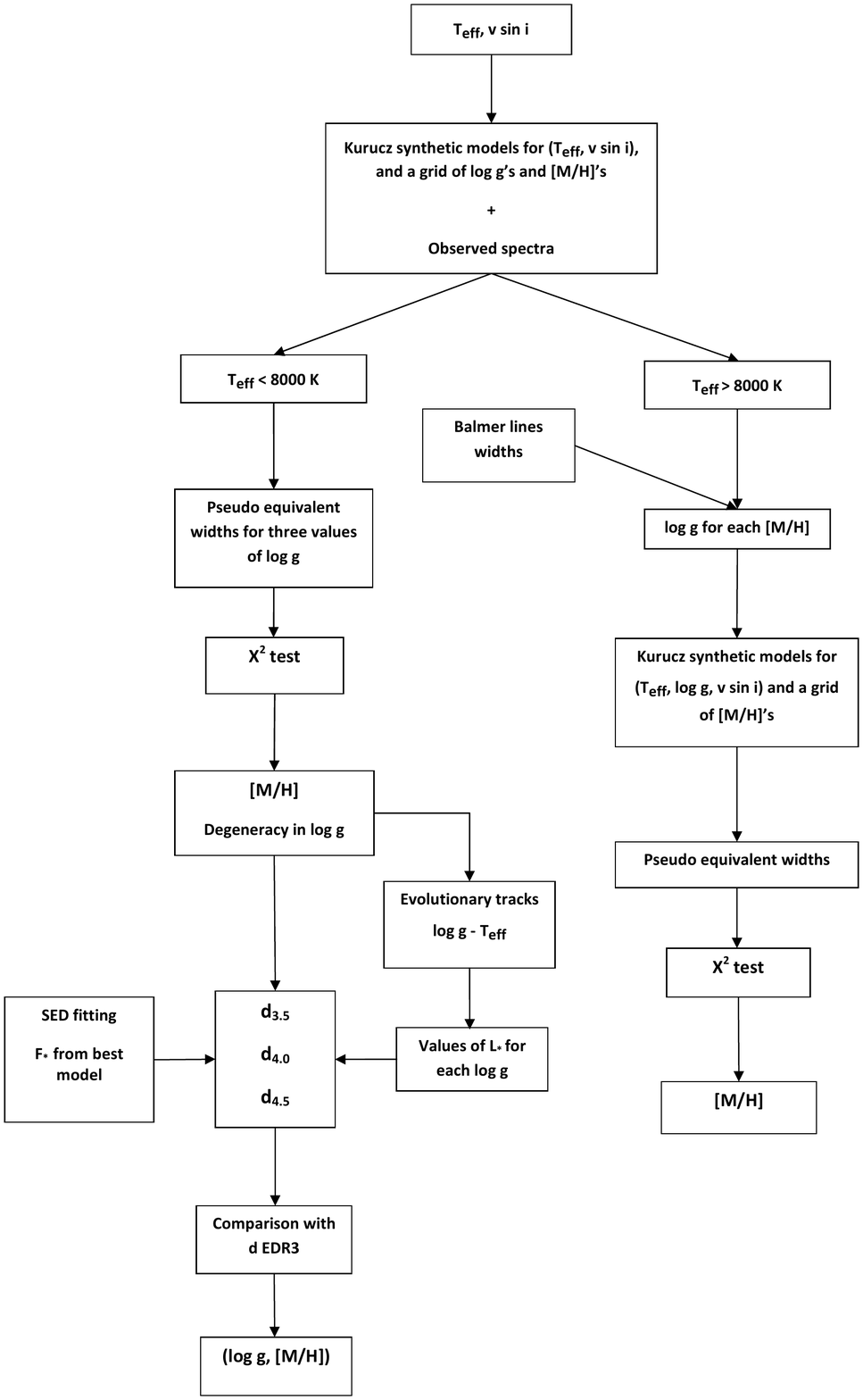}
      \caption{Flow chart summarizing the process followed in the estimation of the $\log g$ and [M/H] values (see Sect. \ref{metallicity}).}
      \label{flow_chart}
\end{figure*}

\begin{figure*} [h]
 \centering
    \includegraphics[width=0.33\textwidth]{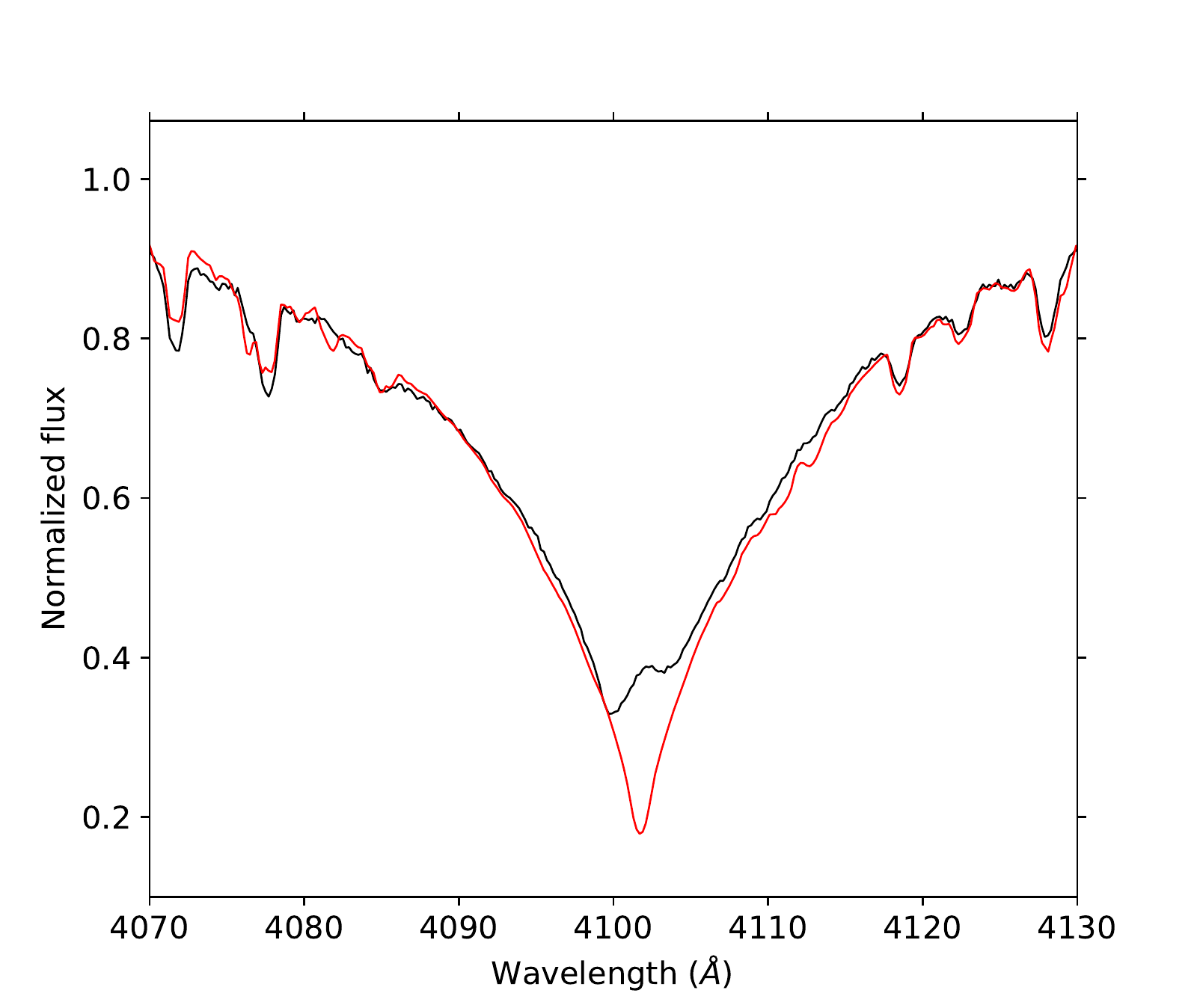}
    \includegraphics[width=0.33\textwidth]{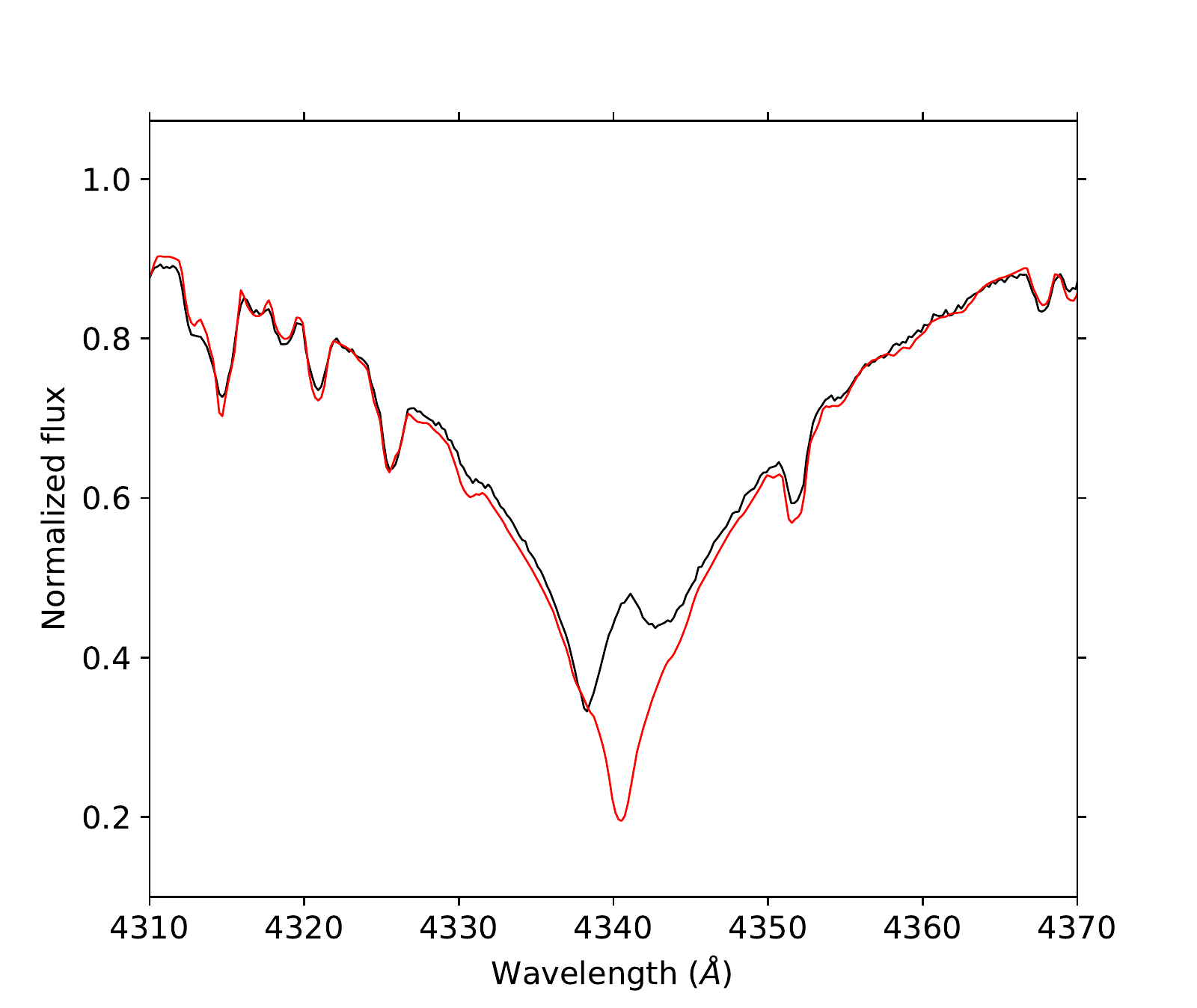}
    \includegraphics[width=0.33\textwidth]{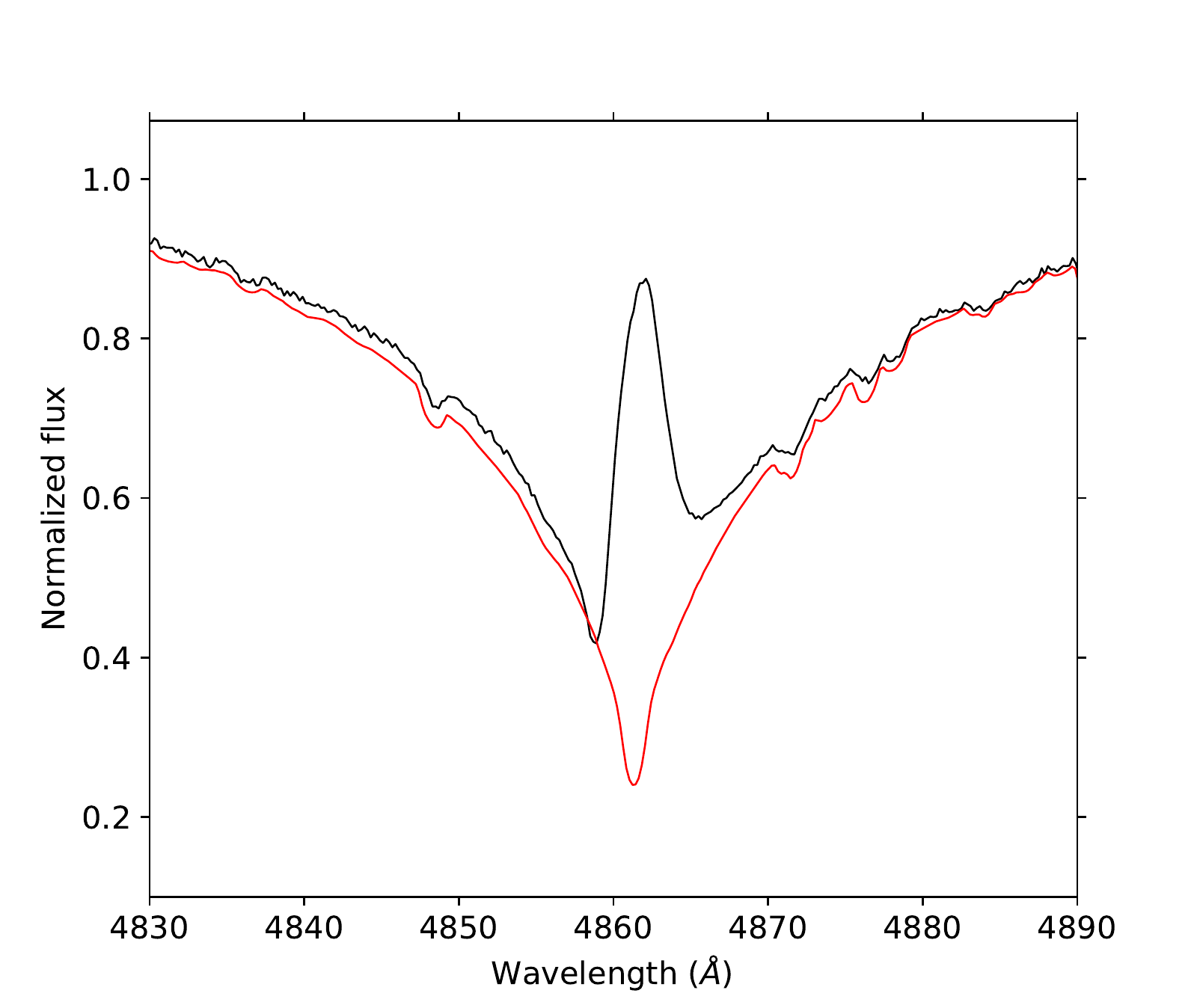}

\end{figure*}

\begin{figure*} [h]
 \centering
    \includegraphics[width=0.33\textwidth]{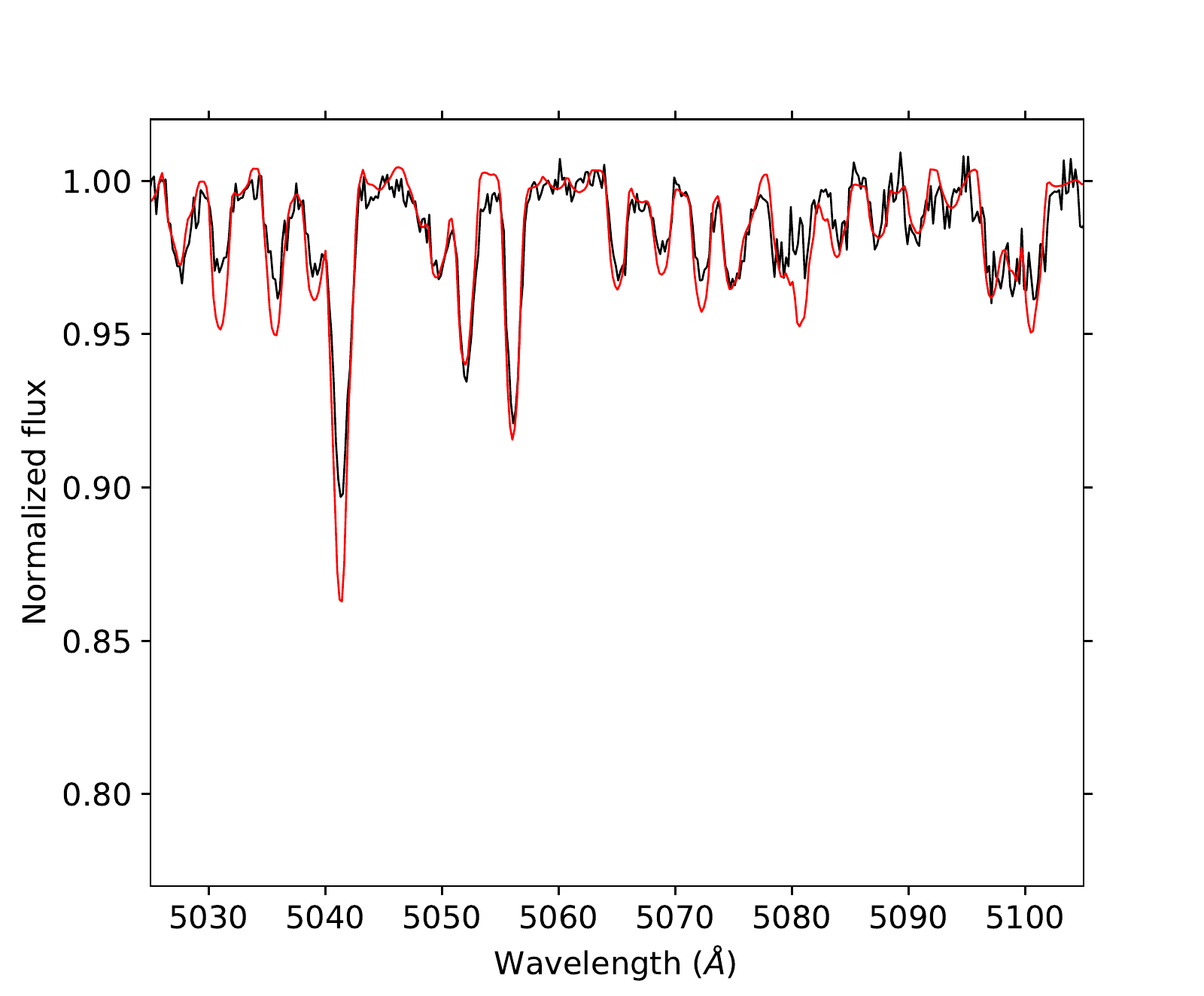}
    \includegraphics[width=0.33\textwidth]{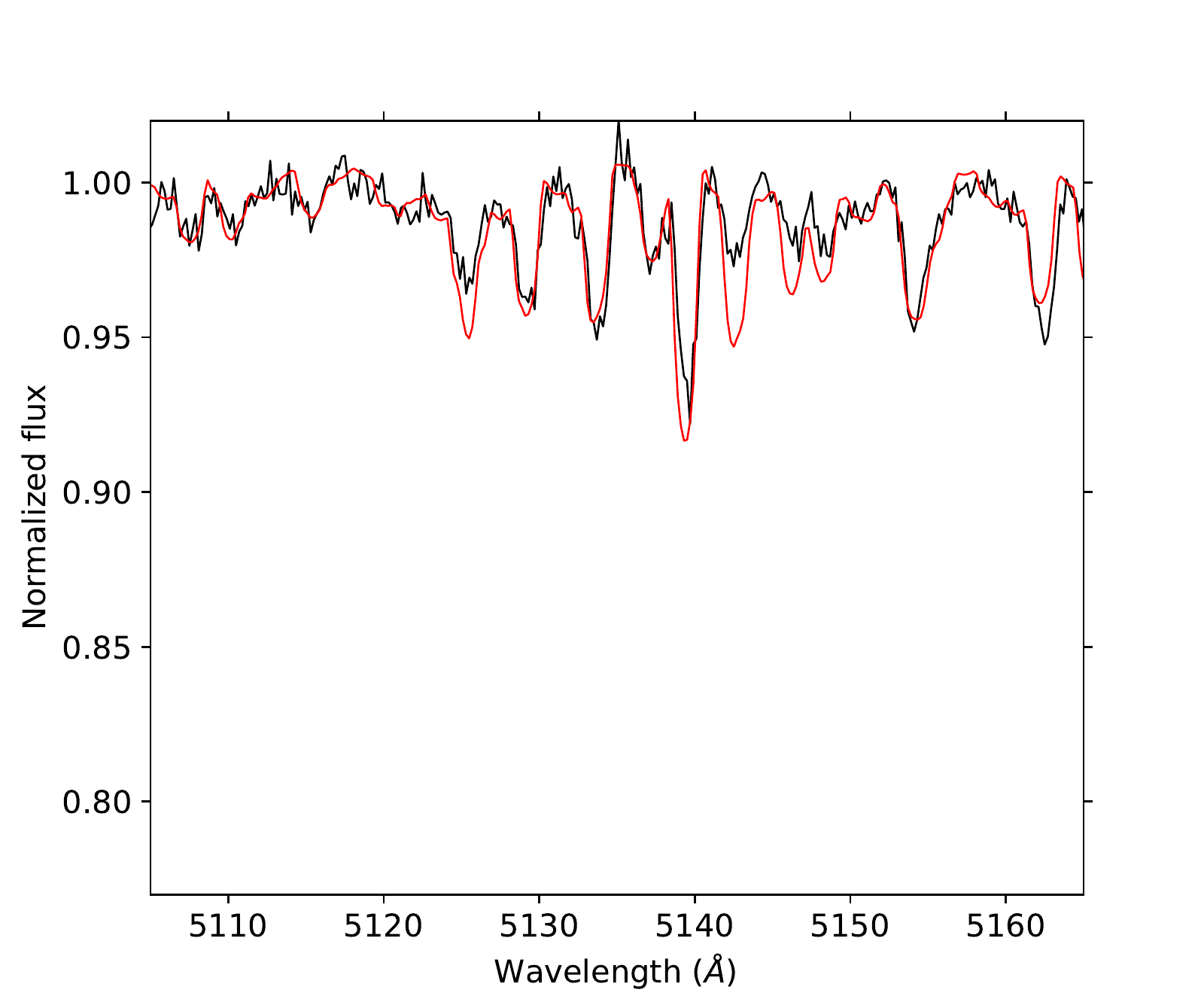}
    \includegraphics[width=0.33\textwidth]{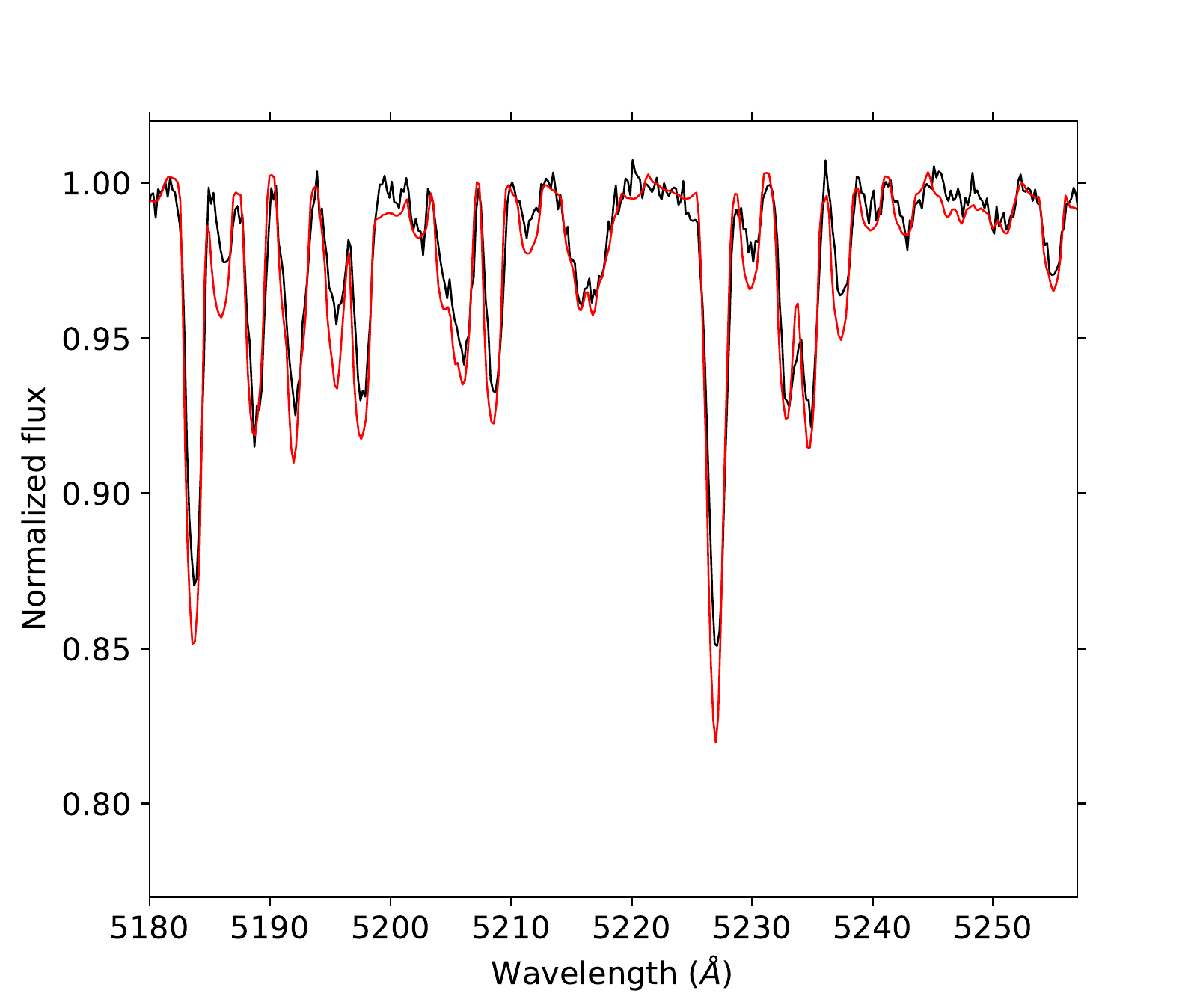}

\end{figure*}

\begin{figure*} [h]
 \centering
    \includegraphics[width=0.33\textwidth]{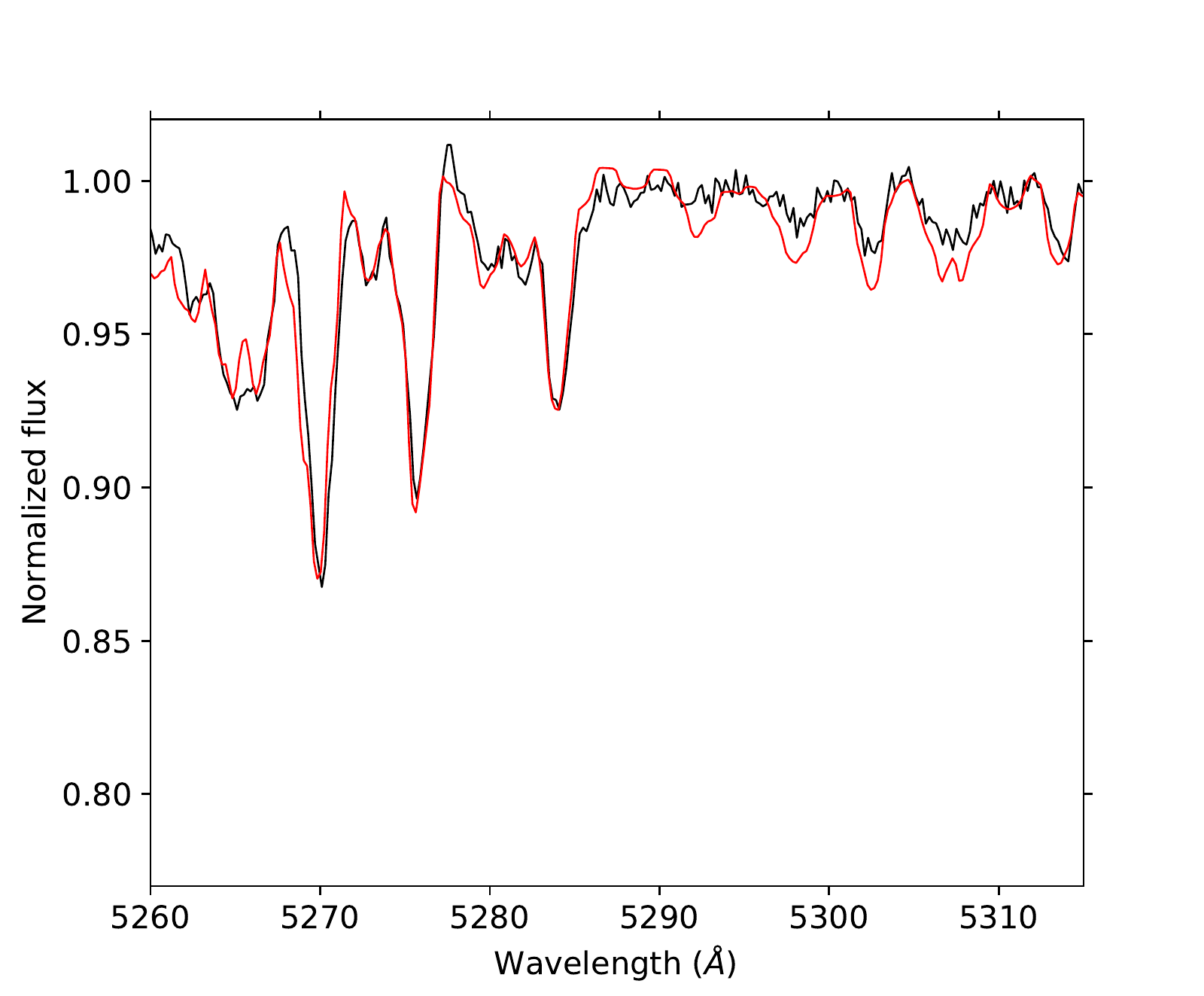}
    \includegraphics[width=0.33\textwidth]{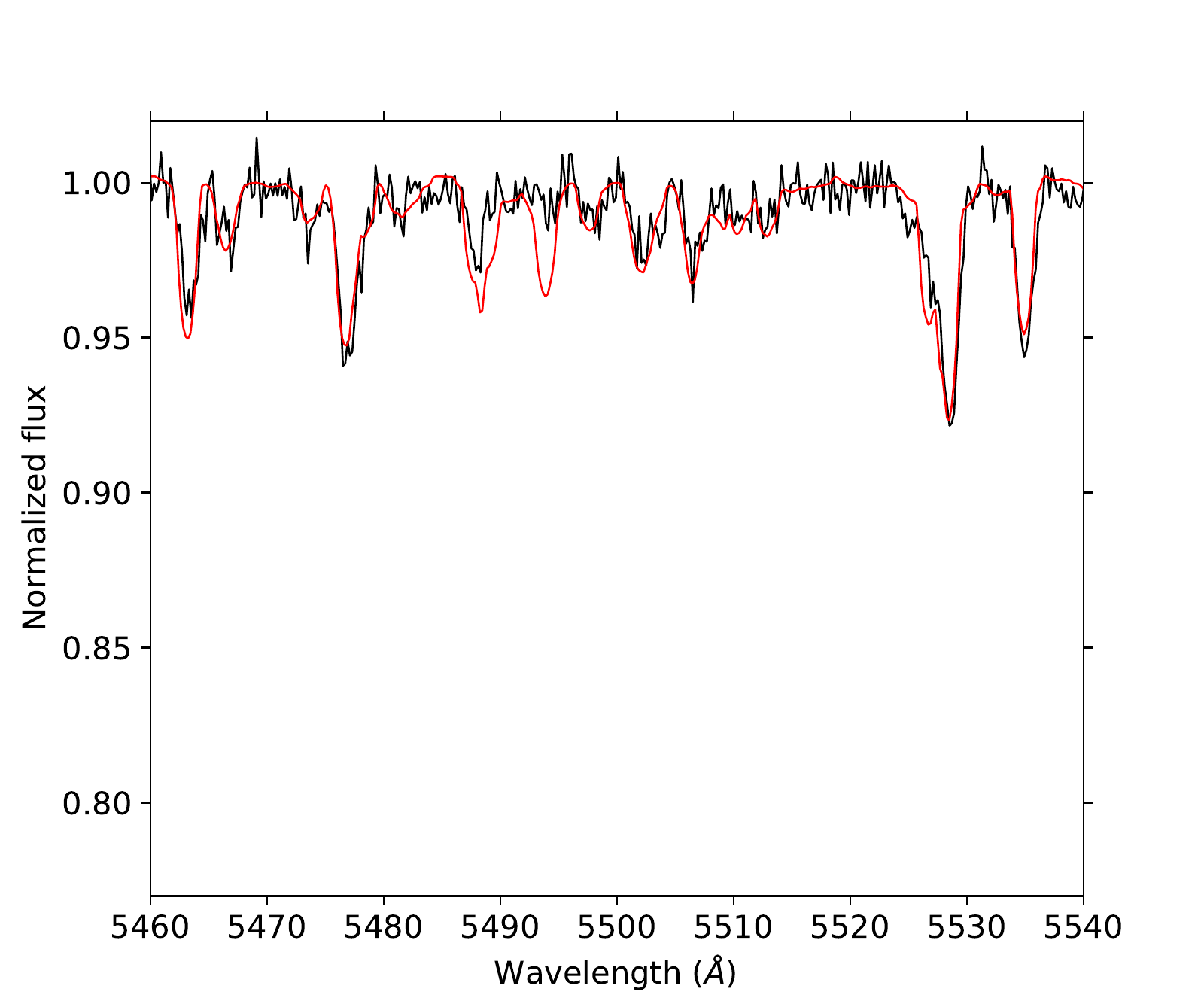}
    \includegraphics[width=0.33\textwidth]{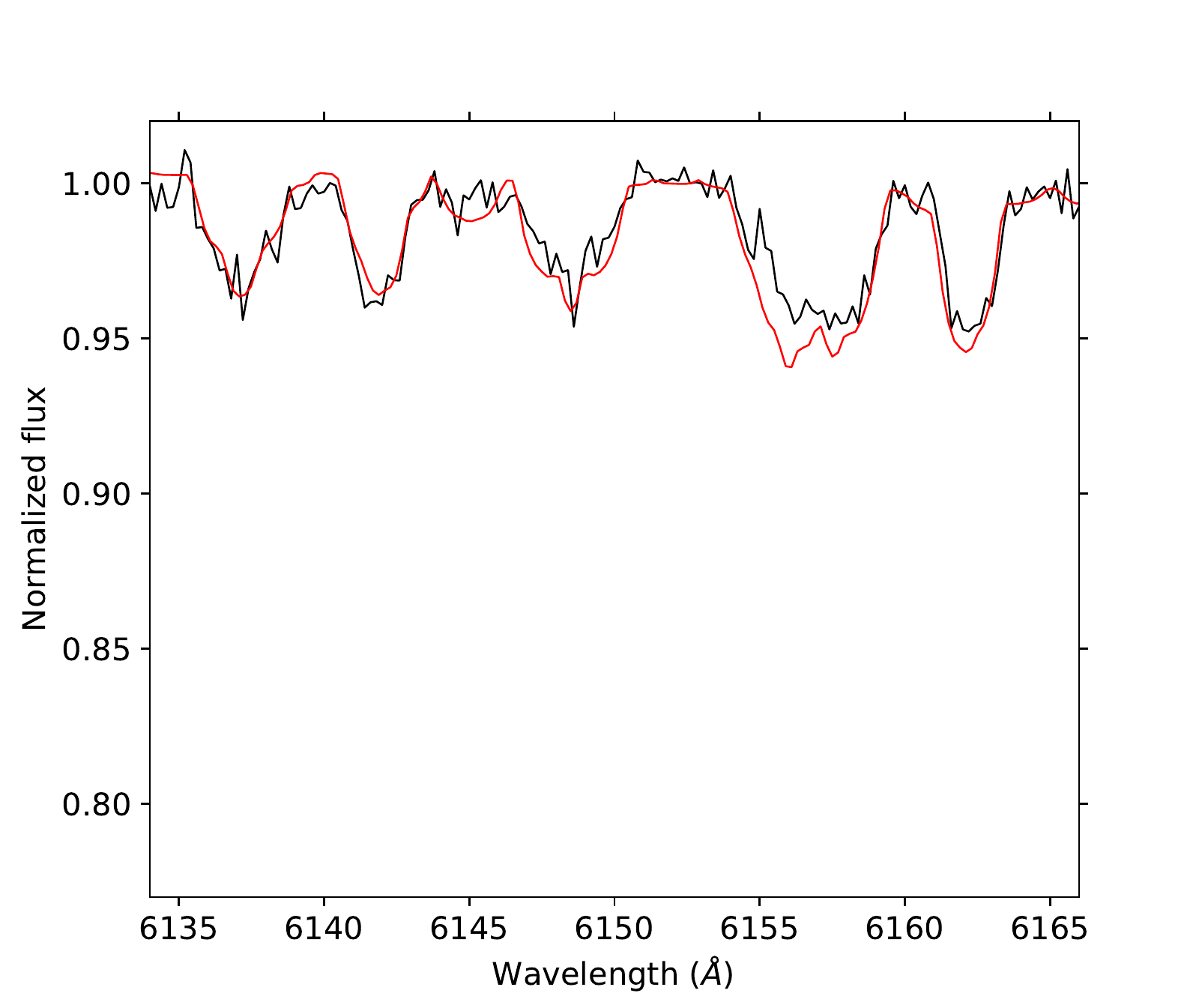}
    \caption{Fits of the Balmer lines H$\delta$, H$\gamma$, H$\beta$, and six regions of the HD 244314 spectrum (black solid line) with the best model obtained in the $\chi^{2}$ test (red solid line). The model has been generated with a T{\tiny eff} = 8500 K, $\log g$ = 4.08, $v \sin i$ = 55 km/s and [M/H] = 0.0. In this example, the line H$\beta$ has not been considered in the estimation of $\log g$ as it shows a strong emission.}
    \label{chi2}

\end{figure*}

 The spectra used in this work
are the final products of the processing of the raw data -and the corresponding calibration files-
through the pipelines developed by the instruments' teams. Special care was taken to normalize the spectra; narrow regions free from lines were chosen in the original spectra and the mean intensities of each small interval were used as clips to build a full continuum using cubic splines. The RASSINE program (\citealt{Cretignier_2020}) has been used in automatic mode to improve the normalization of a few UVES and HARPS spectra. Telluric contamination in our spectra is significant within the ranges $\sim$ 5850-6000 \r{A} and $\sim$ 6450-6600 \r{A}, but these ranges are not used in the analysis (Sect. \ref{metallicity}). Therefore, although telluric correction was not applied this does not affect our results.

%-----------------------------------------------------------------

\section{Results}\label{metallicity}

This section describes how the [M/H] values were obtained from the comparison of the observed spectra with photospheric models, being $v \sin i$ and $\log g$ additional outcomes of the process. Synthetic Castelli-Kurucz spectra computed with the suite of codes {\sc atlas9} \citep{Castelli_2003} were used throughout. These models do not consider disk-to-star accretion, which can produce a continuum excess -and possible line veiling in the coldest sources- mainly at wavelengths shorter than studied in this work \citep[e.g.][]{Muzerolle_2004,Mendigutia_2011a,Mendigutia_2014}. 

\subsection{Projected rotational velocities}\label{vrot}

The first step was to estimate the $v \sin i$ of each star. Although the derivation of [M/H] does not strictly depend on the $v \sin i$ values -rotation does not have an effect on equivalent width, which will be extensively used (see Sect. \ref{M/H})-, that parameter facilitates the visual comparison between observations and models.

Before describing the procedure for estimating $v \sin i$, some words of caution are pertinent. In some cases, when the resolution of the spectrograph is not particularly high -say below 30\,000- that parameter puts some limits to the estimated values of $v \sin i$. The widths of some telluric or arc lines give an idea of the instrumental response of the spectrograph; let us call $\sigma_{\rm inst}$ to that value. Assuming Gaussian profiles, the actual stellar line profiles with widths $\sigma_{\rm real}$, and the instrumental $\sigma_{\rm inst}$ combine according to the expression $\sigma_{\rm obs}^2\!=\!\sigma_{\rm inst}^2\!+\!\sigma_{\rm real}^2$ --where "obs" stands for "observed"--. For the case where $\sigma_{\rm real}\!=\!2\,\sigma_{\rm inst}$, we would have $\sigma_{\rm obs}\!=\!\sqrt{5}\,\sigma_{\rm inst}$. Identifying $\sigma\equiv v$, i.e. widths and velocities, the relative error $(v_{\rm obs}-v_{\rm real})/v_{\rm real}$ would be $\sim\!11$\%. For the particular case of all the XSHOOTER spectra obtained with R=9900, $\sigma_{\rm inst}\!\simeq\!30$ km/s, i.e. values of $v \sin i$ below twice that one would be affected by larger uncertainties and must be taken as upper limits, the situation being more critical for the few stars whose spectra were obtained with even lower resolutions.

The broad, isolated Mg {\sc ii} feature at 4481 \AA{} was generally used to estimate the $v \sin i$ of the objects. For each star, the full width at half depth (FWHD) of that line was varied by changing the $v \sin i$ in the synthetic model with the corresponding T{\tiny eff} from Table \ref{Table1}. The $v \sin i$ that provided a FWHD equal to that of the observed spectra within 5\% was adopted as the final value. Column 4 of Table \ref{Table2} lists the $v \sin i$ values obtained for the sample stars. Alternative photospheric lines were used for the few stars showing a Mg II line with a peculiar profile that makes the estimation of $v \sin i$ challenging. In addition, previous $v \sin i$ values from the literature were taken as a departure point and refined based on the procedure described above for the following stars: CO Ori (\citealt{Herbig_88,Mora_2001,Glebocki_2005}), HBC 217 (\citealt{McGinnis_2018}), HBC 222 (\citealt{McGinnis_2018}), HD 101412 (\citealt{Cowley10}), HD 104237 (\citealt{daSilva_2009}), HD 143006 (\citealt{Jonsson_2020}), HD 169142 (\citealt{Alecian_2013}), PX Vul (\citealt{Pereyra_2009}) and BP Psc (\citealt{Torres_2006}). Finally, the value of $v \sin i$ for BF Ori was directly adopted from \citet{Mora_2001}.

\subsection{Surface gravities and metallicities}\label{M/H}

Figure \ref{flow_chart} illustrates the steps followed in the estimation of $\log g$ and [M/H]. Firstly, for each star we generated an initial grid of Kurucz models with the corresponding T{\tiny eff} and $v \sin i$ listed in Tables \ref{Table1} and \ref{Table2}, $\log g$=3.5, 4.0, 4.5 and [M/H]=$-2.5$, $-2.0$, $-1.5$, $-1.0$, $-0.5$, 0.0, +0.2, +0.5. From this point on, two different paths were followed depending on the T{\tiny eff} of the source.

For stars with T{\tiny eff} $>$ 8000 K, $\log g$ was derived by comparing the observed widths of the Balmer lines H$\delta$, H$\gamma$ and H$\beta$ at intensity 0.80 with those from the models (\citealt{Gray_2009}). The initial grid was interpolated in order to obtain the $\log g$ values that provide the best fit for each line, the final value being the average of the previous. Uncertainties were calculated from the standard deviation of the individual $\log g$ values obtained in each of the considered Balmer lines. Lines affected by strong emission components that do not allow to correctly trace the wings of the absorption profiles were excluded from the analysis (see an example in the top right panel of Fig. 2).

Once $\log g$ was determined, observed and modelled pseudo equivalent widths (pEW) were compared to each other for the different [M/H] values, providing a final value of [M/H] based on a $\chi^{2}$ test. A pEW is defined exactly in the same way as the classical EW, but instead of characterizing a single spectral line, the measurement extends over several lines within a wavelength range $[\lambda_1,\lambda_2]$. The reason for using pEW instead of EW is that the former is less limited by spectral resolution. Indeed, most stars in the sample show large values of $v \sin i$, which makes an individual analysis of the spectral lines difficult due to blending. Moreover, the lower spectral resolution of the XSHOOTER spectra, which are intensively used, prevents us from measuring EWs of individual lines accurately, even in objects with moderately small values of $v \sin i$. Although in a different context -G and K giants-, a similar method to estimate metallicities was developed by \citet{Gray_2002} \citep[and see e.g.][for a practical application of that formalism]{Lillo_Box_2014}. 

The wavelength region used to calculate the pEWs was [5000, 7000] \r{A}. The windows initially selected were 80-\r{A} wide, excluding the broad interval containing H$\alpha$, and the number of such windows per object was $\sim$ 6. However, for each source a careful selection of what intervals inside these windows were usable was done. Several factors prevented us from selecting the same regions homogeneously for all the stars, namely: i) the presence of emission lines; ii) the normalization of particular sections of the stellar spectrum; iii) a low SNR, which causes some lines not to be discernible  and difficulties to trace the continuum; iv) scarcity of spectral lines (especially in  very hot stars); v) circumstellar contribution in absorption, superimposed to the photospheric lines; and vi) variability. Regarding the last point, when multi-epoch spectra are available for a given source, they were checked to avoid regions where variability in the profiles of the metallic lines was apparent. It should also be mentioned that some  spectra had to be renormalized manually in order to better estimate [M/H].

Using as a reference for the final [M/H] value the one given by the $\chi^{2}$ test, the corresponding uncertainty was estimated from the individual [M/H] best-fitting value of each spectral region considered. An average of the absolute errors was computed, assuming a minimum uncertainty of 0.10 dex.

A different procedure was followed for stars with T{\tiny eff} $\leq$ 8000 K. The widths of the Balmer lines are insensitive to changes in surface gravity for such sources (\citealt{Gray_2009}); spectral indicators for this kind of objects based on lines ratios are difficult to apply given their typically large values of $v \sin i$.  Therefore, in these cases moved directly to the step where the pEWs are calculated in the observed spectrum and in the whole collection of models (Fig. \ref{flow_chart}). Hence, the result of the $\chi^{2}$ test yields a value of [M/H], but with a degeneracy in $\log g$. In fact, the three best models with $\log g$=3.5, 4.0, 4.5 fit the regions of the observed spectrum practically in the same way, hardly showing any variation in the pEWs of the regions explored. In order to find out which $\log g$ is closer to the stellar value, we have used an indirect method that consists of determining which of these $\log g$ is consistent with the distance from Gaia (E)DR3 \citep[][see also Table B.1 of \citealt{Guzman-Diaz_2021}]{Lindegren20, Gaia_2022}. 

Let us denote d$_{3.5}$,  d$_{4.0}$ and  d$_{4.5}$, the distances implied from the three values of $\log g$. These distances are derived from the expression $d\!=\!\sqrt{L_{*}/4\pi F_{*}}$, where L$_{\ast}$ is the stellar luminosity, and F$_{\ast}$ is the total observed flux from the star. The fluxes were calculated using the VOSA\footnote{http://svo2.cab.inta-csic.es/theory/vosa} tool by integrating the -dereddened- Kurucz ODFNEW/NOVER models (\citealt{Castelli_1997})  computed for T{\tiny eff}, [M/H] from the $\chi^{2}$ test, and the three values of $\log g$. Although the visual extinction (A$_{v}$) was left virtually free in the fits, their final values obtained for each log g are similar to those shown in Table B.1 of \citet{Guzman-Diaz_2021}. The stellar luminosities were derived by translating  the points (T{\tiny eff}, $\log g$) (with $\log g$=3.5, 4.0, 4.5) from the  T{\tiny eff} - $\log g$ HR diagram to the T{\tiny eff} - $\log L_{*}$ diagram. PARSEC V2.1s\footnote{https://people.sissa.it/~sbressan/parsec.html} evolutionary tracks (\citealt{Bressan_2012}) were used. Each one of the three points (T{\tiny eff}, $\log g$) has a one-to-one corresponding point (T{\tiny eff}, $\log L_{*}$); once d$_{3.5}$,  d$_{4.0}$ and  d$_{4.5}$ are computed, the stellar gravity is the one whose derived distance matches that from Gaia (E)DR3. Interpolation between distances was used when required. A typical error of 0.05 dex on $\log g$ was adopted.

\begin{figure}[h]
   \centering
   \includegraphics[width=\hsize]{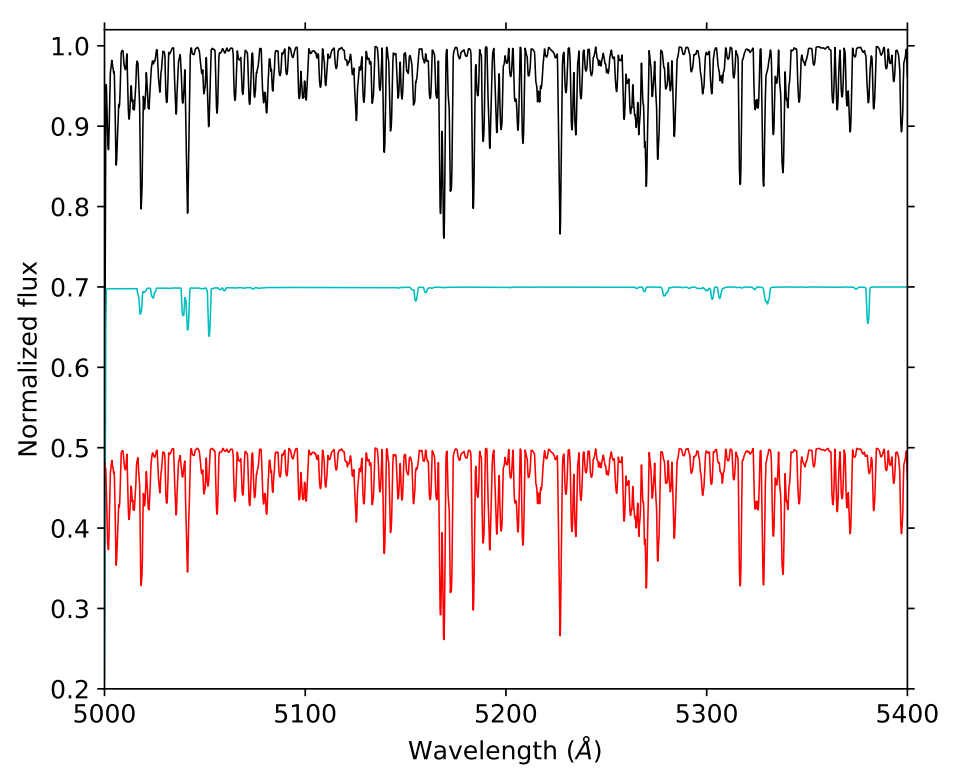}
      \caption{A model with T{\tiny eff} = 8000 K, $v \sin i$ = 50 km/s and [M/H] = 0.0 in the 5000-5400 \r{A} region is represented in black, while only the C, O, and S lines generated with the same parameters are shown in cyan. The result of subtracting the volatile elements from the first model is shown in red. The models were shifted on the y-axis for clarity.}
      \label{sanity_check}
\end{figure}

\begin{figure}[h!]
   \centering
   \includegraphics[width=7.3cm]{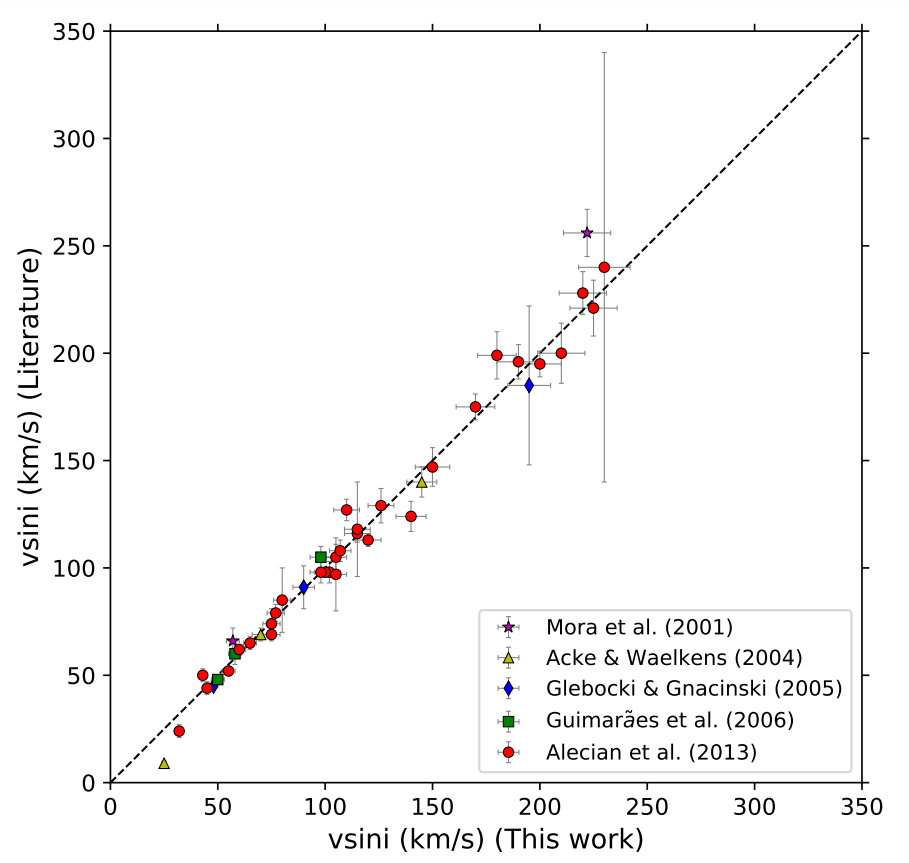}
   \includegraphics[width=7.3cm]{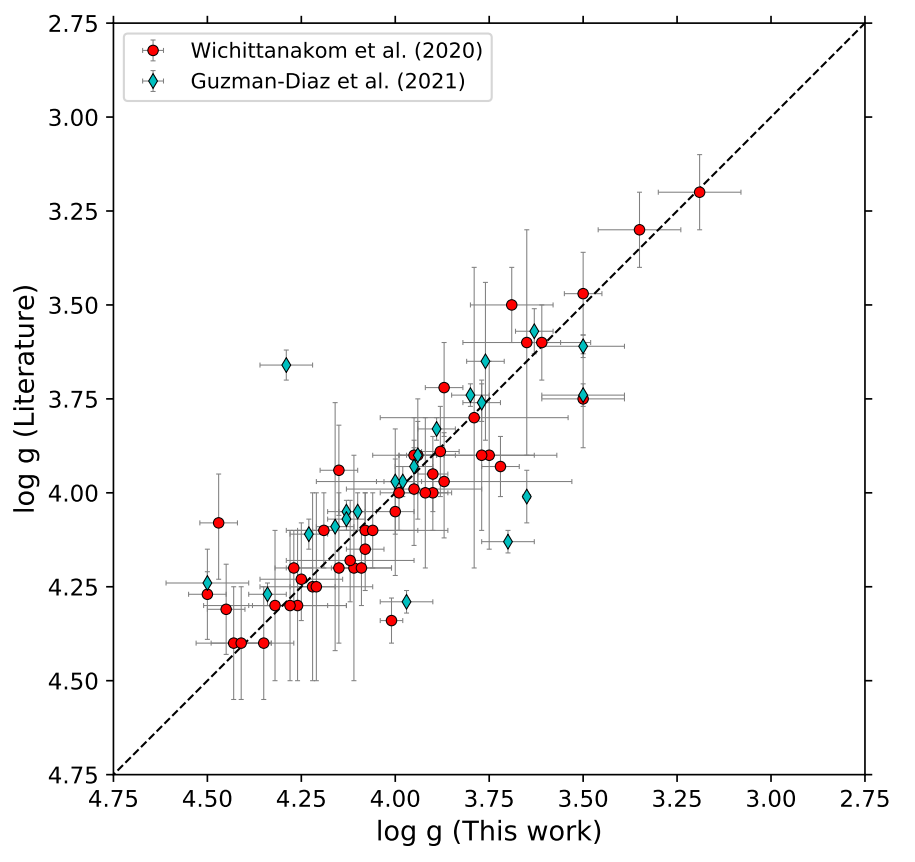}
   \includegraphics[width=7.3cm]{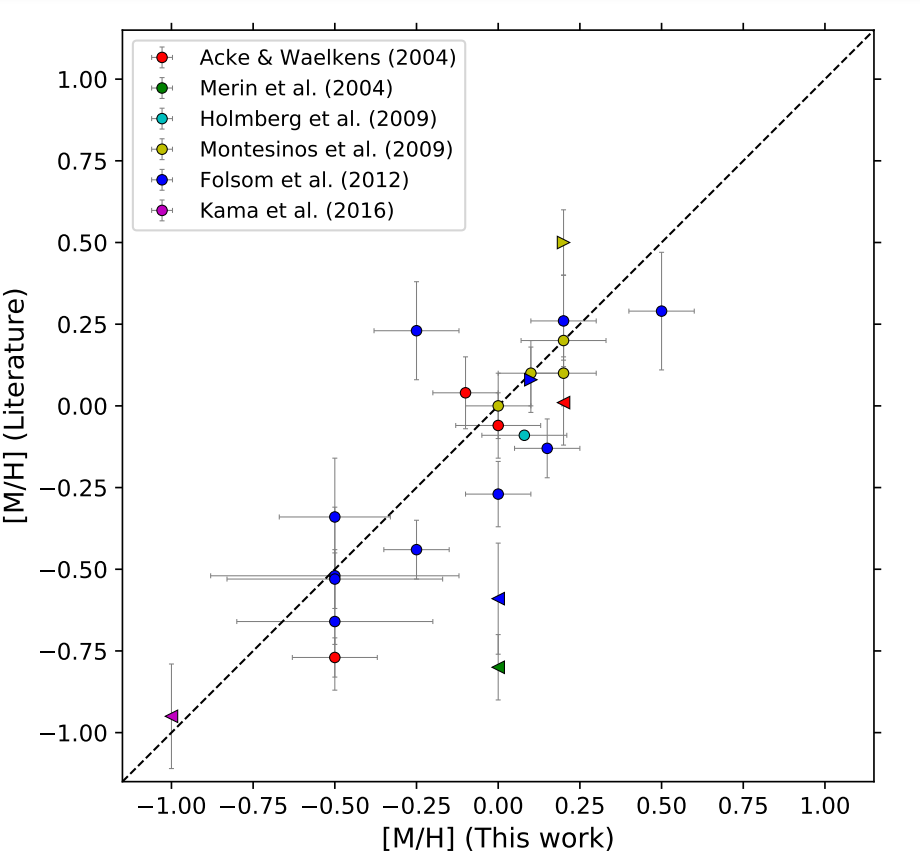}
      \caption{The $v \sin i$, $\log g$ and [M/H] are compared with the values found in the literature (as indicated in the legends). Error bars represent the uncertainties in these parameters, while the lower and upper limits are denoted by right- and left-pointing triangles, respectively. The black dashed line indicates equal values.}
      \label{metal_comp}
\end{figure}

\begin{figure*}
 \centering
    \includegraphics[width=0.33\textwidth]{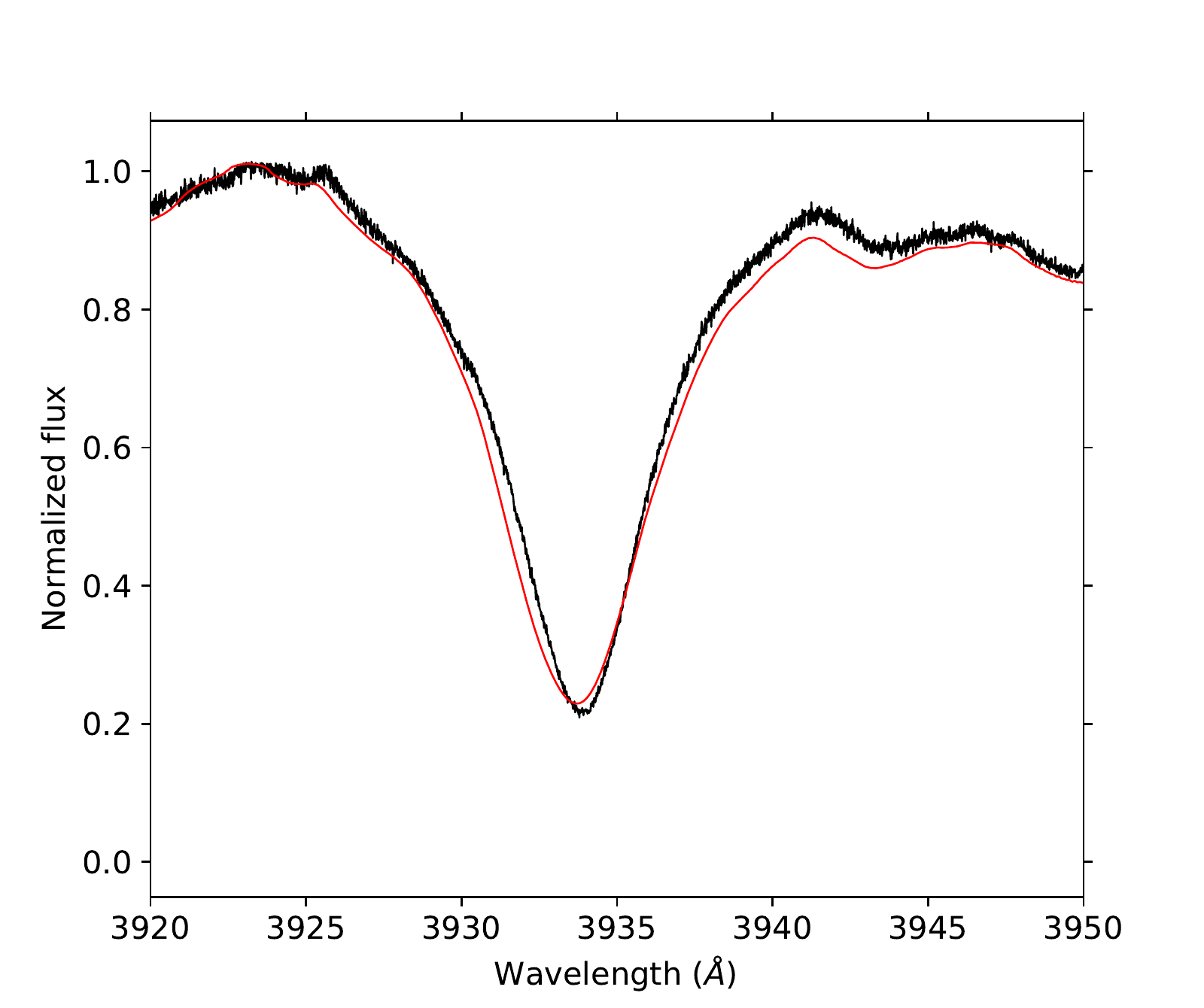}
    \includegraphics[width=0.33\textwidth]{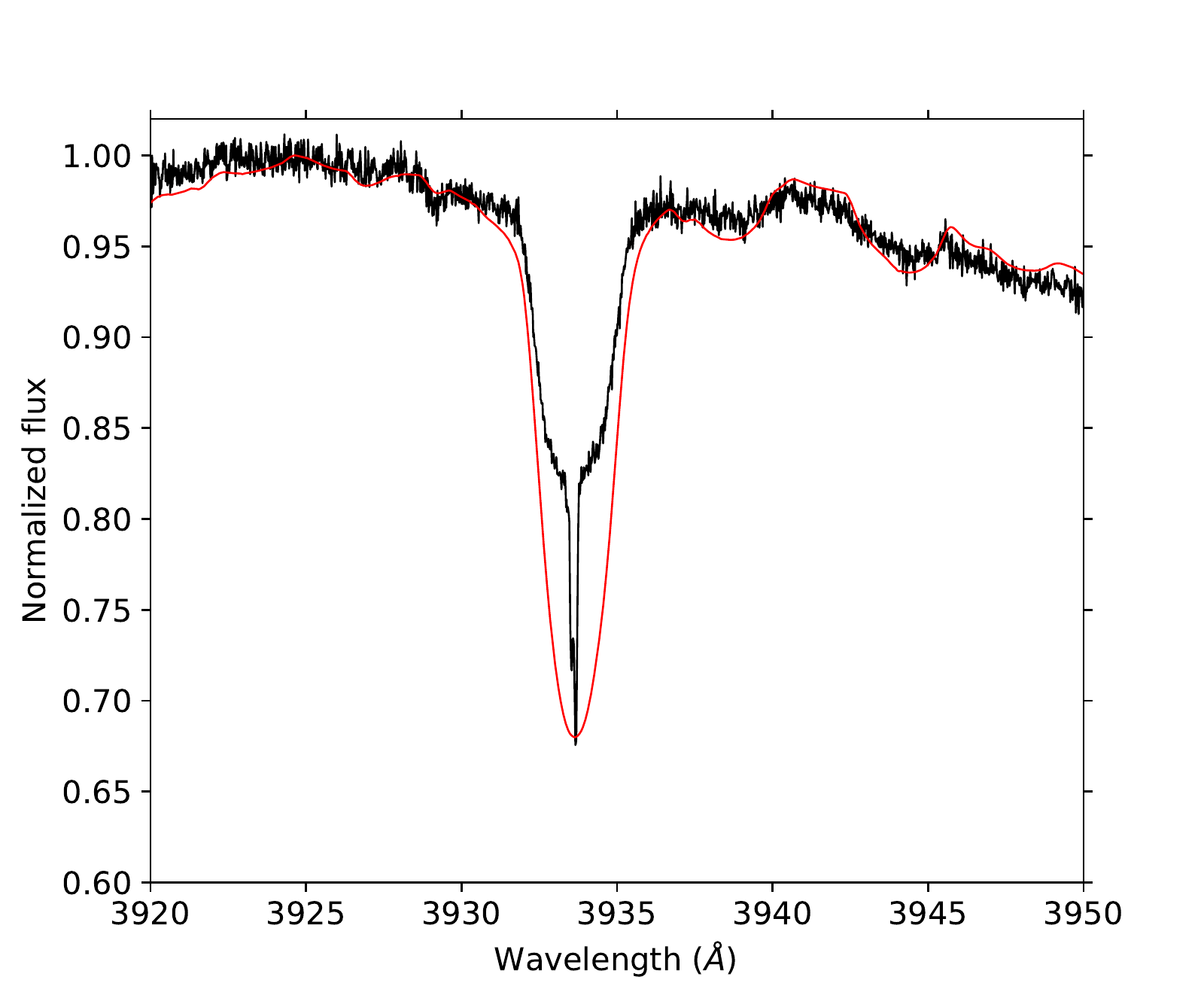}
    \includegraphics[width=0.33\textwidth]{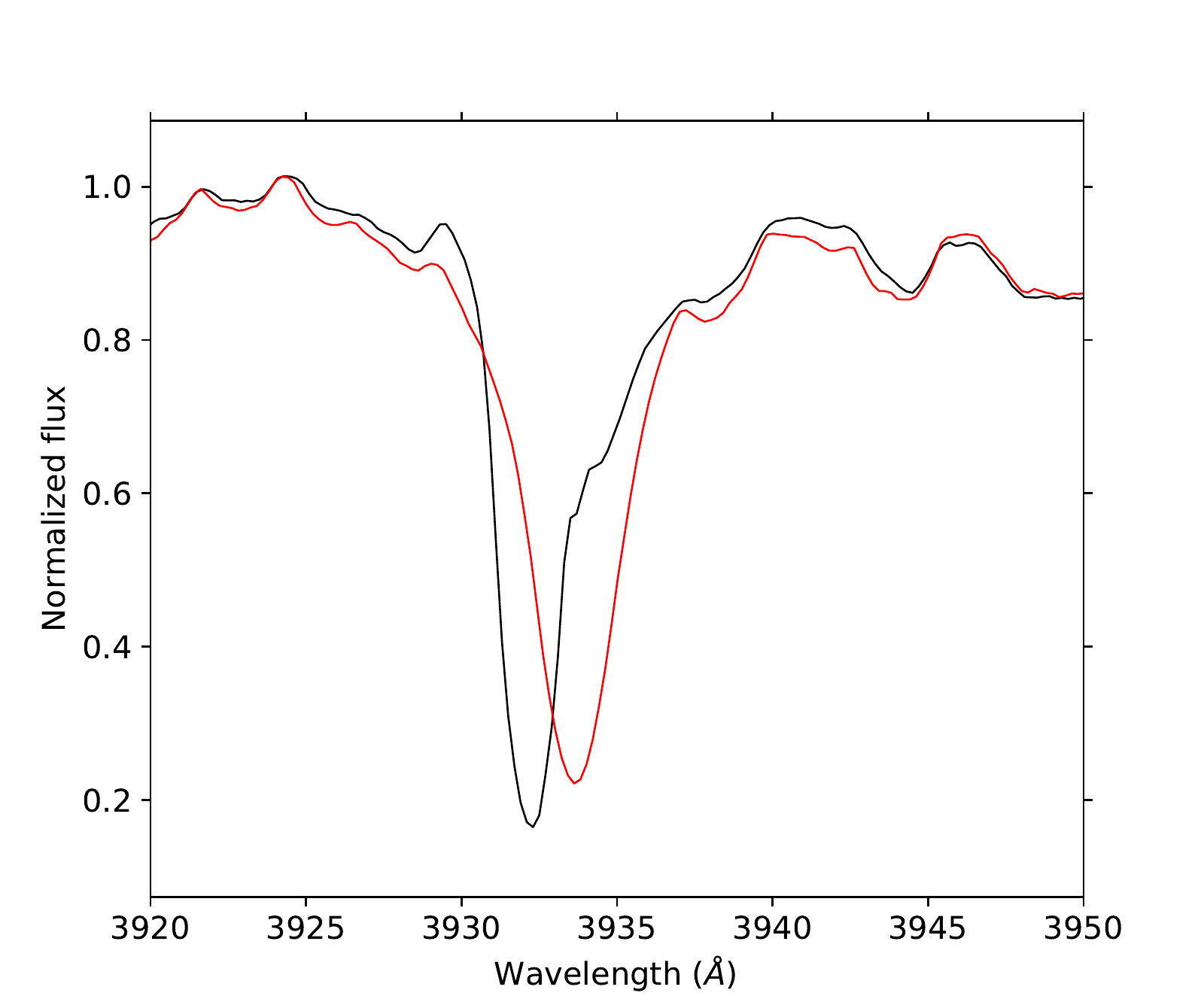}
    \caption{Comparison of the Ca {\sc ii} K lines of the spectra of the stars (black solid lines) HD 39014 (left), HD 132947 (middle) and HD 31648 (right), with those of the best models (red solid lines). A good fit is only achieved in the first case, where no circumstellar contribution or peculiar profiles are apparent.}
    \label{CaIIK}
\end{figure*}

In some stars, two different values of [M/H] provide similar values of $\chi^{2}$. For these cases, we generated new models by interpolating the values of [M/H] and repeated the $\chi^{2}$ test in order to check if a better fit was obtained. Should this occur, we assumed this new value; otherwise, the [M/H] with the lowest $\chi^{2}$ retrieved from the previous test was kept.

Table \ref{Table2} lists the final values inferred for $\log g$ and [M/H] and their corresponding errors in Cols. 5 and 6. Figure \ref{chi2} shows a representative example of the best fit obtained applying the procedure described above for HD 244314.

\subsection{Interpretation of the [M/H] derived in this work}

In order to interpret the results, it is important to emphasize explicitly that the method used does not take into account variations of individual abundances of specific elements, i.e. for a given [M/H], all abundances are scaled in the same manner to the solar one by a factor $10^{\rm [M/H]}$, the whole procedure providing an average metal abundance for each star.

Since the hypothesis by \citet{Kama15} linked a deficit of refractory elements to the presence of dust traps and cavities potentially caused by Jovian planets, we carried out the following exercise: All lines of C, O, N, and S in the interval 5000-7000 \r{A}
were identified and the pEWs were computed in the same regions both in the stellar spectra and in the models, avoiding
the small windows where lines of those volatile elements appear. Given that the number of lines of volatiles is much less than that of refractories, it turned out that the new [M/H] values do not show significant changes when compared with the initial calculations.

The previous test was done for six stars of the sample with different values of T{\tiny eff}, $v \sin i$ and [M/H] to cover a representative set of parameters, finding no significant differences with respect to the values already listed in Table \ref{Table2}. Figure \ref{sanity_check} shows an example of how the subtraction of the C, O, N and S lines from a synthetic model hardly produces  any change in the 5000-5400 \r{A} region, which serves to visualize why the [M/H] measurements are mainly determined by the abundances of refractory elements regardless of the abundances of volatiles.
%-----------------------------------------------------------------

\section{Analysis and discussion}\label{results}

\subsection{Consistency tests}
\label{consistency}

\subsubsection{Comparison with previous results}\label{comp}

The top panel of Fig. \ref{metal_comp} compares the values estimated here for $v \sin i$ with previous values from the literature, which are available for 63$\%$ of the stars in our sample. For those values in the literature where no associated error was found, an uncertainty of 5\% was assumed. Results are in good agreement for the majority of the stars, being the mean relative error $\sim$ 7\%. The previous analysis excludes the 10 stars in Table \ref{Table2} with $v \sin i$ values based or directly adopted from the literature (Sect. \ref{vrot}). 

Concerning $\log g$, our values and previous ones based on the literature are plotted against each other in the middle panel of Fig. \ref{metal_comp}. In this case the comparison has been possible for the whole sample. $\log g$ values not available in \citet{Wichittanakom20} have been computed from the expression $g=GM_{*}/R_{*}^{2}$, taking the stellar mass, M$_{*}$, and radius, R$_{*}$, from \citet{Guzman-Diaz_2021}. The corresponding uncertainties have been calculated by error propagation. The absolute error for most objects is below 0.20 dex, with 18\% of them showing an error larger than that value.

Finally, the comparison between the [M/H] values from this work and from the literature is included in the bottom panel of Fig. \ref{metal_comp}, which has been possible for 36$\%$ of the stars in the sample. Literature values come from \cite{Montesinos09} and \citet{Kama15}. In turn, \citet{Kama15} compiled individual abundances of certain elements from \citet{AckeWaelkens_2004} and \citet{Folsom_2012}, except for HD 34282
(\citealt{Merin_2004}), HD 100546 (\citealt{Kama_2016}) and HD 142527 (\citealt{Holmberg_2009}). In order to convert such abundances into metallicitiy values, the Appendix of \citet{Montesinos09} has been followed, considering only the abundances of Fe, Si and Mg given in \citet{Kama15}. There is a reasonable agreement between our results and the literature ones, and the sources identified as having low- and high-[M/H] coincide. The mean absolute difference is $\sim$ $\pm$ 0.15 dex, excluding the sources with lower and upper limits. Notes on specific sources where the differences between our [M/H] values and those from the literature are relevant can be consulted in Appendix \ref{appB}. 

\subsubsection{The Ca {\sc ii} K line }\label{line_Ca}

The depth of the Ca {\sc ii} K line is used in main sequence A-type stars as the prime estimator of
their effective temperature (\citealt{Gray_2009}). However, following other works
devoted to the determination of stellar parameters of HAeBes \citep[e.g.][]{AckeWaelkens_2004,Folsom_2012}, this line has also been excluded from the whole procedure devised in this work. The reason is that most stars in the sample show Ca {\sc ii} K lines with circumstellar contributions and/or peculiar profiles that pure photospheric models are not able to reproduce. In addition, the blue part of the spectrum would be affected by veiling in those cases where the accretion contribution to the total flux starts to be comparable to the photospheric flux. 

With the previous caveat in mind, we have checked how the best-fitting models from the $\chi^{2}$ tests match the observed Ca {\sc ii} K profile for stars with T{\tiny eff} between 8000 K and 10500 K (A5 - B9). For the few stars with normal, symmetric profiles, the models reproduce the line profile with a reasonable level of accuracy, which constitutes an additional proof of self-consistency. Figure \ref{CaIIK} shows three representative examples. For HD 39014 the model fits the Ca {\sc ii} K line reasonably well, HD 132947 shows a circumstellar component superimposed to the photospheric profile, and finally, HD 31648. presents a peculiar profile that the model is unable to reproduce. 

\begin{figure}[h!]
   \centering
   \includegraphics[width=\hsize]{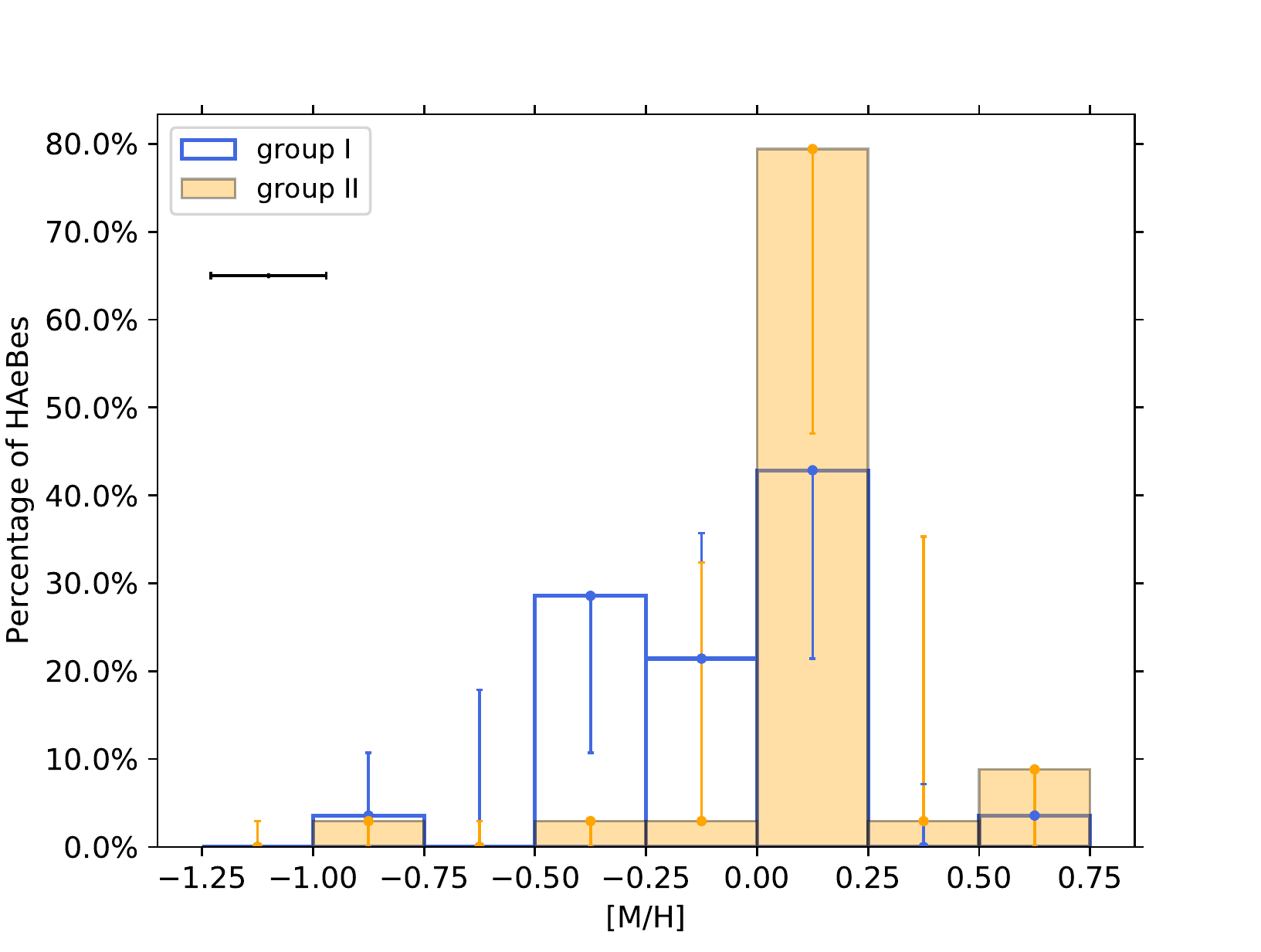}
      \caption{Histograms comparing the distributions of metallicities and SED classification, as indicated in the x-axis and the legend. Vertical error bars represent the variations of the number of stars contained in each bin considering the individual uncertainties in [M/H], which typical value is represented by the horizontal error bar below the legend.}
      \label{histograms_Meeus}
\end{figure}

\begin{figure}[h!]
   \centering
   \includegraphics[width=\hsize]{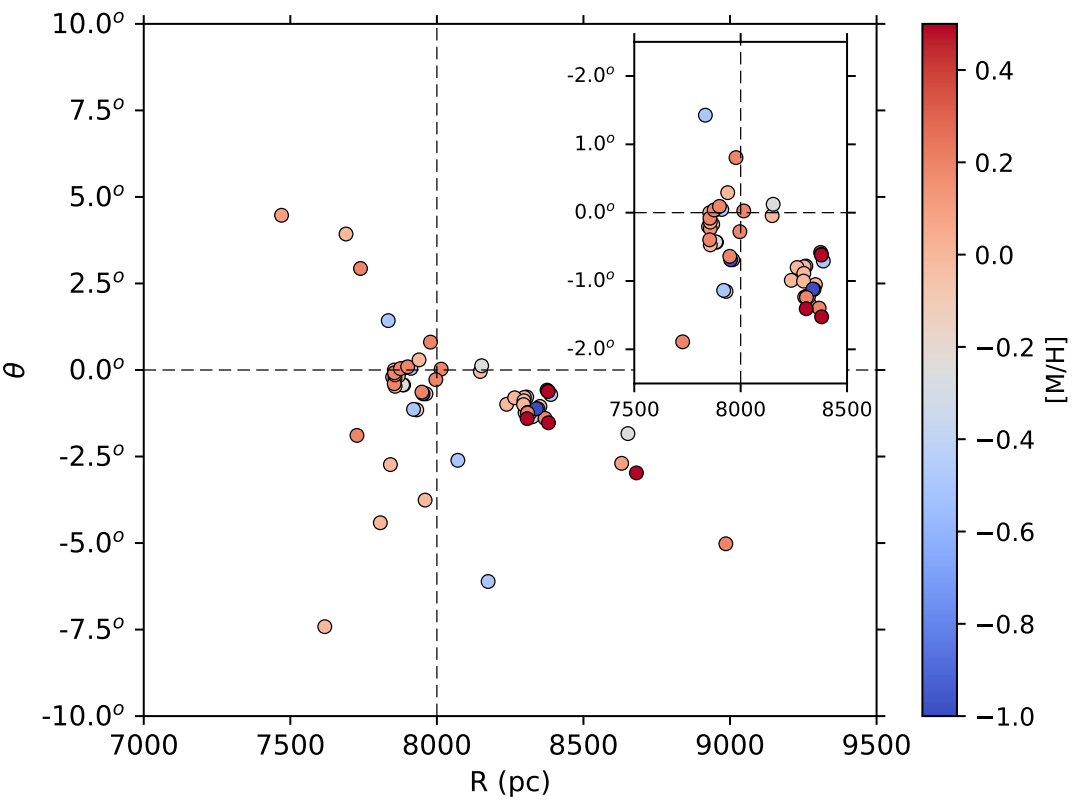}
      \caption{Two-dimensional representation of the distribution in the Galaxy of the HAeBes studied in this work. The Galactic azimuthal angle in the y-axis is plotted against the distance to the Galactic center in the x-axis. The colour bar indicates their [M/H] values. The most crowded region of the plot has been zoomed in. The values for the sun -with [M/H] = 0- are also indicated with the vertical and horizontal dashed lines, for reference.}
      \label{theta_radius}
\end{figure}

\subsection{[M/H] and SED groups}\label{Kama}

Figure \ref{histograms_Meeus} shows the [M/H] distribution of the sample analyzed, where the blue and orange histograms refer to group I and group II sources, respectively. Although most HAeBes tend to show [M/H] values close to solar, the relative fraction of group I sources having sub-solar values is larger ($\sim$ 55\%) than that for group II ($\sim$ 10\%). In particular, the median [M/H] is -0.10 for group I and +0.14 for group II HAeBes. We carried out a Kolmogorov-Smirnov (KS) test to verify whether the group I and group II HAeBes  have different [M/H] distributions or not. The result obtained is that there is a negligible probability (p-value = 0.003) that both samples are drawn from the same parent distribution at a 1\% significant level. An Anderson-Darling (AD) test, more sensitive to the tails of the distributions than the KS test, also rejects the null hypothesis that both group I and group II samples are drawn from the same parent distribution at the same significance level. Whether the upper and lower limits in the histograms are considered or not does not significantly alter the results provided by the KS and AD tests.  

In short, our enlarged sample of HAeBes with homogeneously derived [M/H] values and SED classification according to the \citet{Meeus_2001} scheme statistically confirms the finding by \citet{Kama15}: group I and group II sources tend to show different distributions of [M/H] values, with group I sources being less metallic than group II. 

\begin{figure}[h]
 \centering
    \includegraphics[width=\hsize]{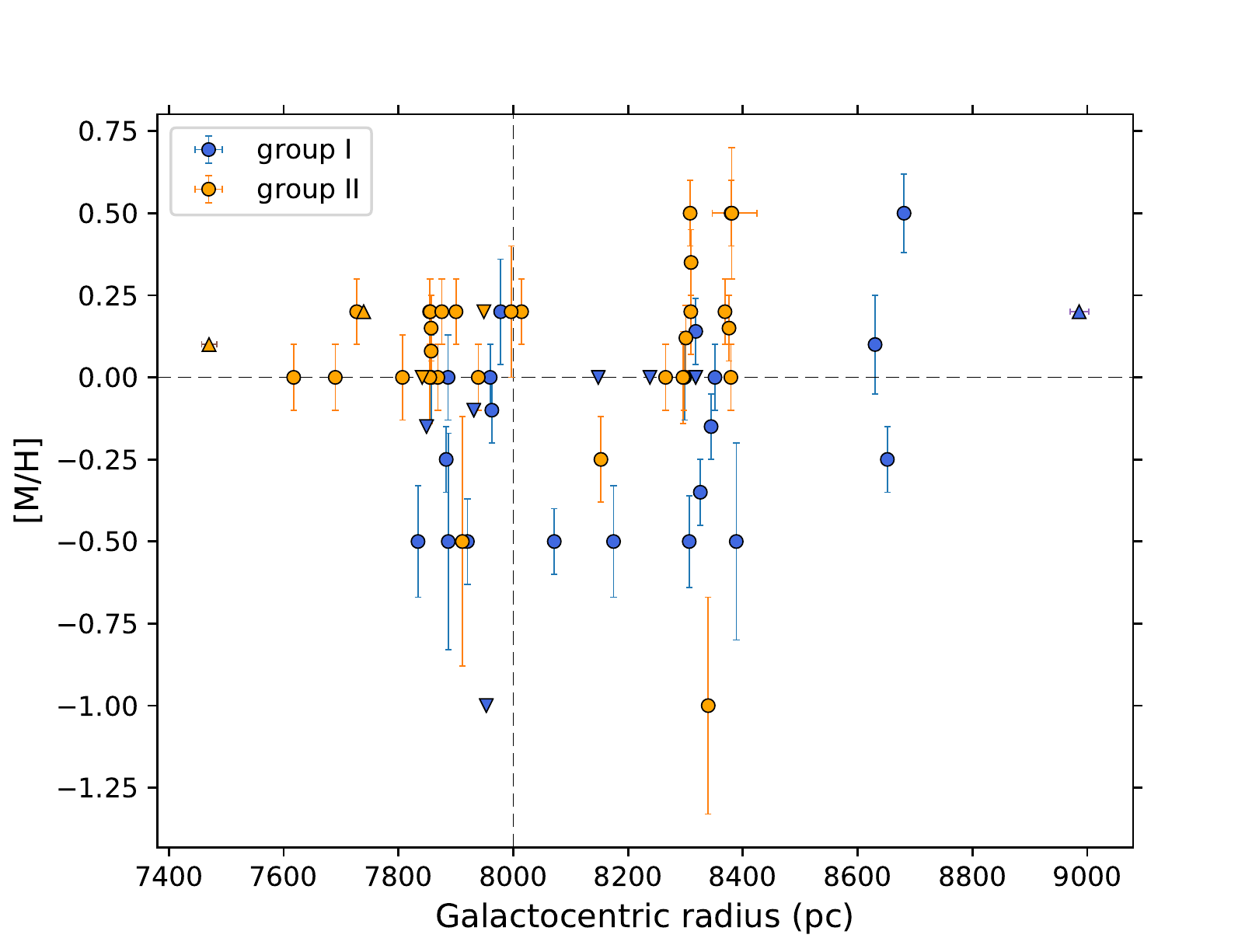}
    \includegraphics[width=\hsize]{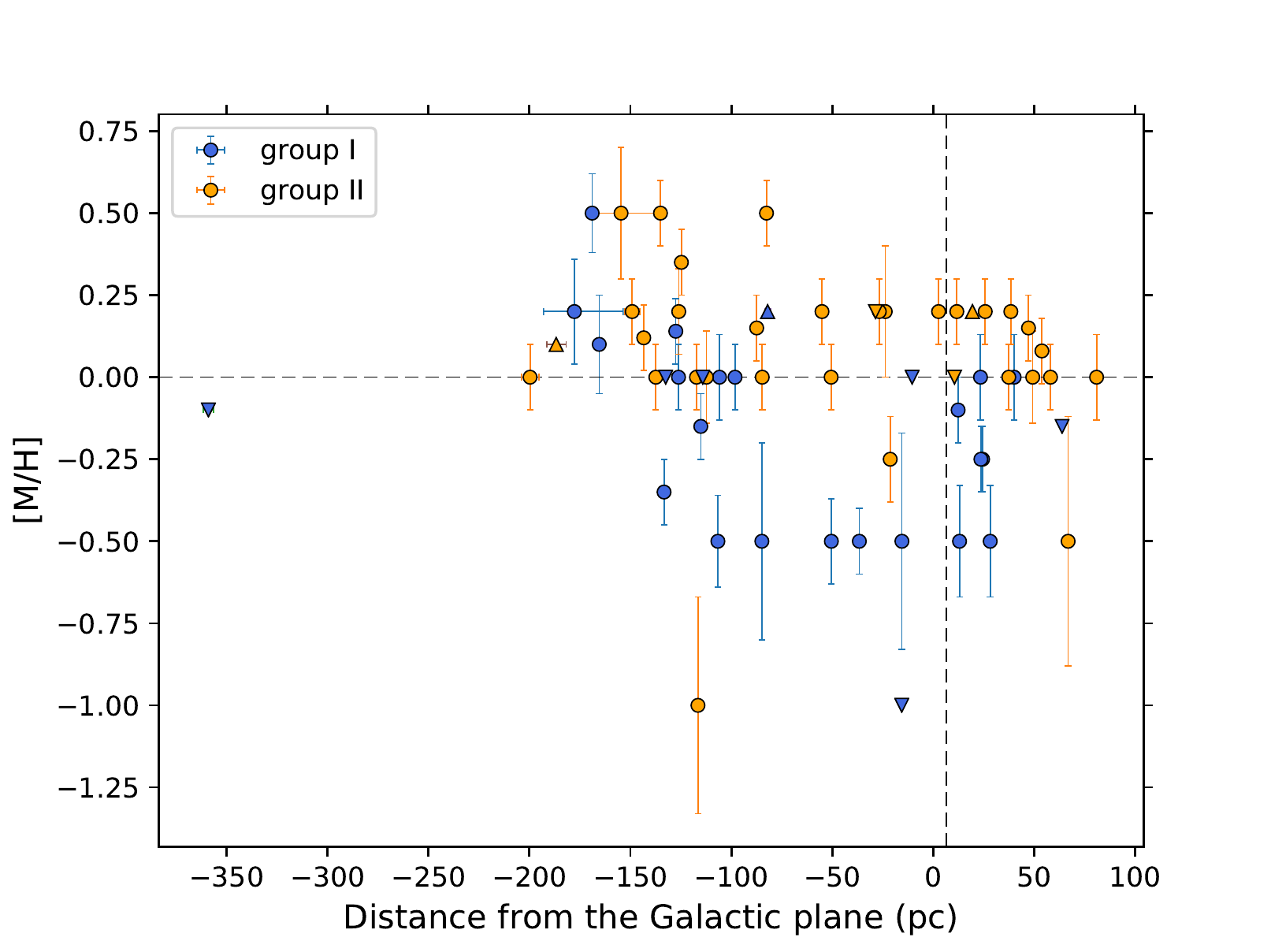}
    \caption{Metallicity of the stars in our sample versus the distance to the Galactic center (top) and to the galactic plane (bottom). Group I and II sources are color-coded as indicated in the legends. The lower and upper limits are indicated by the up and down triangles. The values of the sun are also indicated with the vertical and horizontal dashed lines, for reference.}
    \label{Galac}
\end{figure}

\subsection{A link with planet formation?}\label{interpretation}

In the following sections we aim to interpret the previous observational trend linking low stellar metallicities with group I sources, focusing on the plausibility of the scenario in which giant planets trapping refractory elements are more frequent around HAeBes with such a type of SED \citep{Kama15} 

\begin{figure}[h!]
   \centering
   \includegraphics[width=\hsize]{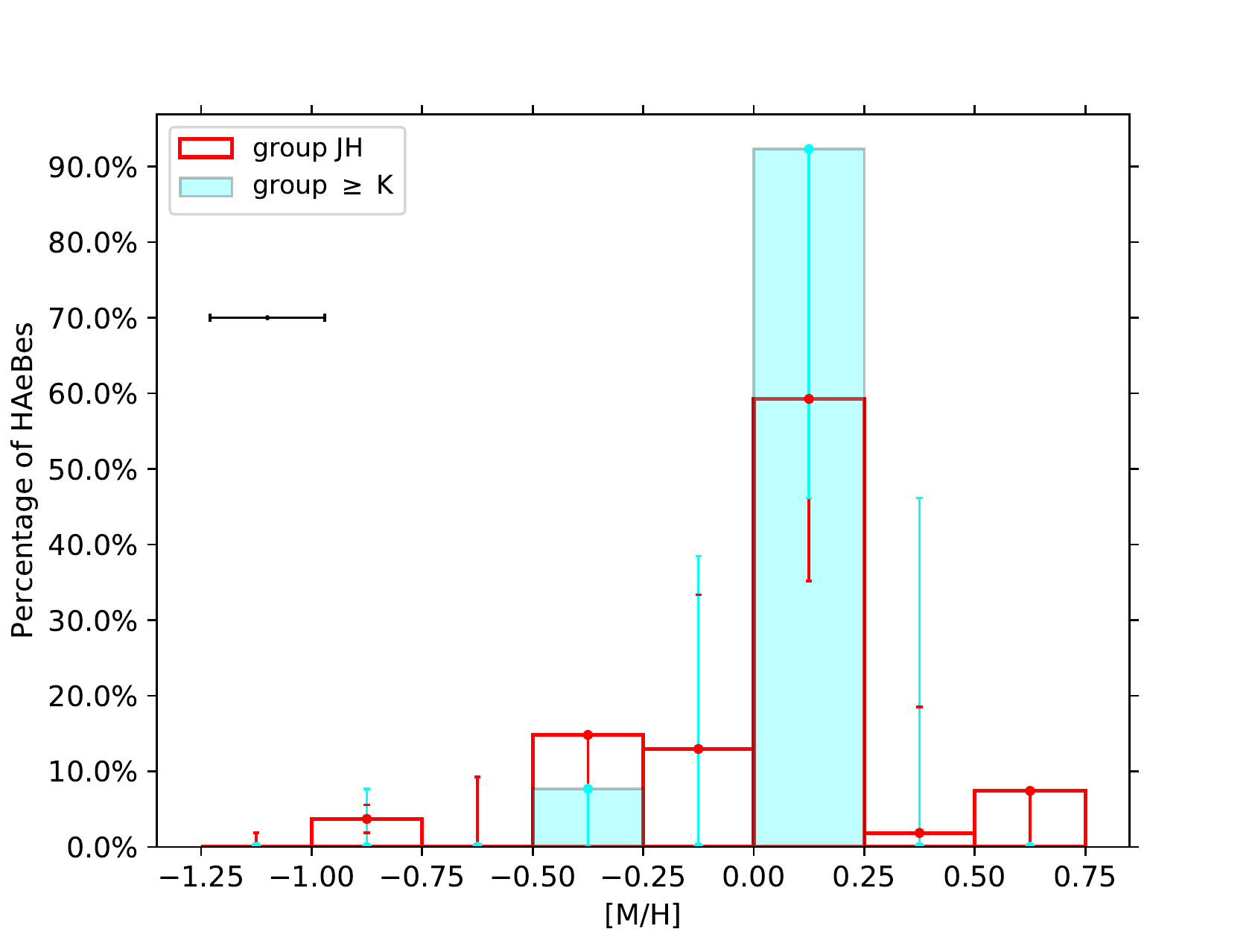}
      \caption{Histograms comparing the distributions of metallicities and JHK classification, as indicated in the x-axis and the legend. Vertical error bars represent the variations of the number of stars contained in each bin considering the individual uncertainties in [M/H], which typical value is represented by the horizontal error bar below the legend.}
      \label{histograms_JHK}
\end{figure}

\subsubsection{Galactic distribution of group I and group II HAeBes}\label{distribution}

The first question to be addressed is whether different stellar metallicities are actually associated to different SED groups or those could instead be driven by environmental effects. Given that the mixing timescale for radiative sub-photospheric regions like the ones of HAeBes is similar to their typical ages ($\sim$ Myr), that possibility cannot be neglected. For instance, one may wonder if the low- and high-[M/H] stars are located in different clusters or regions within the Galaxy, in which case the stellar metallicities may reflect the local, initial conditions. However, no significant difference is found concerning the clustering properties of the low- and high-[M/H] stars in our sample, which appear mixed together and sharing common regions (Fig. \ref{theta_radius}). A similar result is obtained for the whole sample classified by \citet{Guzman-Diaz_2021} in 112 group I and 70 group II HAeBes.

We also tested if the trend linking lower metallicities to group I sources could be related to the well-known empirical relations showing that stellar abundances depend on both the distance to the Galactic center and to the Galactic plane \citep[e.g][and references therein]{Adibekyan2014,Hawkins_2022}. The top and bottom panels of Fig. \ref{Galac} show the [M/H] estimated in this work versus the galactocentric radius and the vertical distance to the galactic plane, respectively. The scatter observed in both panels indicates that the observed differences between the typical metallicities of group I and group II sources are not driven by [M/H] gradients on a Galactic scale, at least for the HAeBes analyzed here.

We can therefore conclude that the [M/H] values of the stars in our sample do not depend on where they were born, keeping the hypothesis by \citet{Kama15} as a possible cause for the different metallicities between group I and group II HAeBes.

\begin{figure}[h]
 \centering
    \includegraphics[width=\hsize]{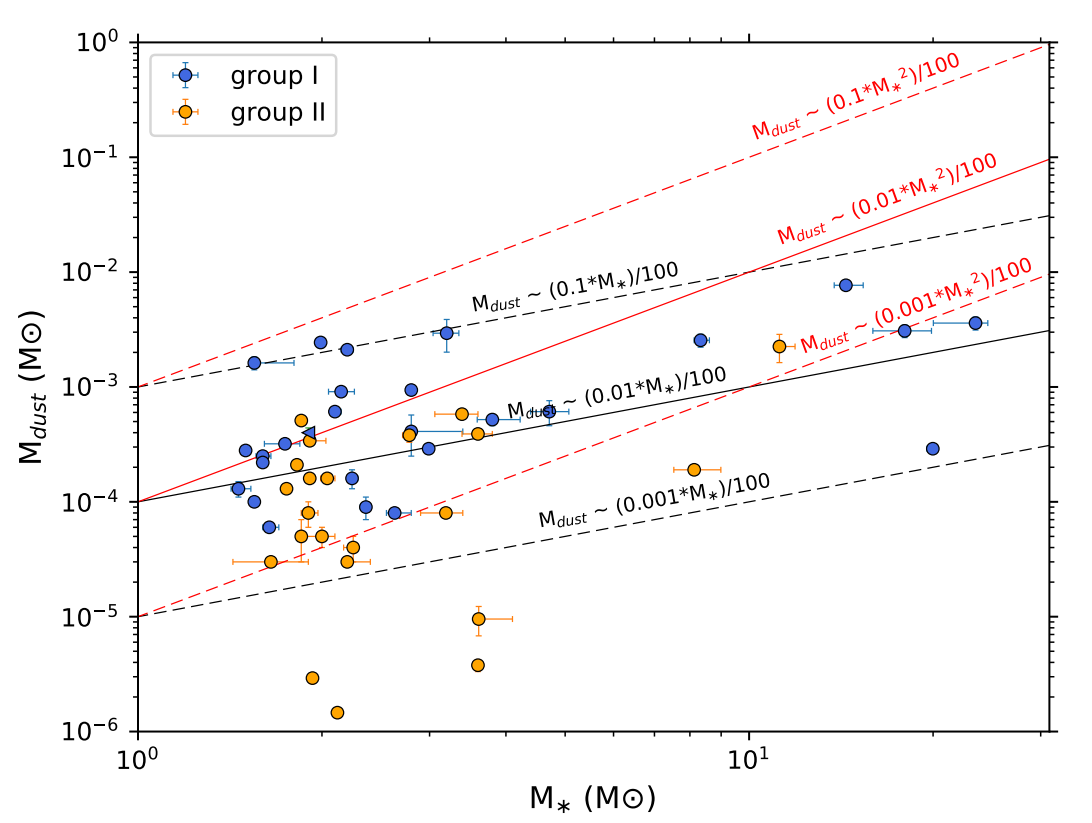}
    \caption{Dust disk mass versus stellar mass for most HAeBes having (sub)mm emission as compiled in \citet{Guzman-Diaz_2021}. Group I and II sources are color-coded as indicated in the legend. Typical power-law trends with exponents 1 and 2 are also shown in black and red, respectively. The corresponding $\pm$ 1 dex scatter are indicated with dashed lines.}
    \label{stellar_disk_mass}
\end{figure}

\subsubsection{[M/H] and transitional disks}\label{JHK}

The last column of Table \ref{Table1} lists the type of SED shown by each star in the sample according to the "JHK classification" carried out in \citet{Guzman-Diaz_2021}. According to this, stars are divided in four groups J, H, Ks, and $>$ Ks, indicating the shortest near-infrared band where excess emission over the photosphere is already present. As detailed in \citet{Guzman-Diaz_2021} (and references therein), such a classification can be related to the size of the holes in the inner dust disks, being the Ks and $>$ Ks sources associated to the so-called "transitional disks" with relatively large inner holes.

Figure \ref{histograms_JHK} shows the [M/H] distribution of the sample analyzed, where this time the red and cyan histograms refer to non-transitional and transitional disks as inferred from the JHK classification. Based on the KS and AD tests, the null hypothesis that both sub-samples are drawn from the same parent distribution cannot be rejected, contrasting with the analogous analysis in Sect. \ref{Kama} concerning the SED groups. 

That [M/H] is not related to the presence of transitional disks -as inferred from SEDs- indicates that the stellar abundances are not generally affected by the different processes that may cause the appearance of dust inner holes. Such holes can be related to other mechanisms apart from the presence of planets, like grain growth, dust settling or photoevaporation \citep[e.g.][and references therein]{Espaillat14}. Indeed, most transitional disks in HAeBes seem to be related with the latter process and the high stellar luminosities, being much more frequent for stars with the earliest spectral types (HBes) than for the rest \citep[see the discussion in][and references therein]{Guzman-Diaz_2021}. In contrast, the cavities associated to group I sources may be related to the potential presence of giant planets, which could cause the metal depletion of the central stars \citep{Kama15}.  

%It should be noted that the sample of stars belonging to group K is rather small (13 stars). This is because the sample of stars in this work comprises up to late B spectral types. As described in \citet{Guzman-Diaz_2021}, objects within that group tend to have higher luminosities and masses, corresponding to earlier B-type stars.

\subsubsection{Dust disk masses of group I and group II HAeBes}\label{disk_mass}

The formation of giant planets requires a large enough mass reservoir in the circumstellar environment. In fact, models predict better chances to host giant planets for the sources having larger disk masses, M$_{\rm disk}$, within a similar stellar mass range \citep[e.g.][]{Alexander2007, Andrews_2013}. \citet{Guzman-Diaz_2021} found that the stellar mass distributions of hundreds of group I and II HAeBes are similar, providing M$_{\rm disk}$ based on (sub)mm continuum emission for many of them. Figure \ref{stellar_disk_mass} shows the stellar versus dust disk mass (M$_{\rm dust}$ = M$_{\rm gas}$/100 $\simeq$ M$_{\rm disk}$/100) distribution for most of such sources. The stars from the sample of \citet{Guzman-Diaz_2021} with upper and lower limits in the continuum fluxes were discarded. The ones having excessive continuum emission leading to M$_{\rm dust}$ $>$ 0.01 M$_{\odot}$, probably due to contamination from the surroundings, were not included in the plot either. It is apparent from the 27 group I and 22 group II HAeBes plotted in Fig. \ref{stellar_disk_mass} that although both groups span over the same stellar mass range, M$_{\rm dust}$ --and hence M$_{\rm disk}$-- for group I stars tend to be larger than those for group II. Indeed, the median M$_{\rm dust}$ for group I and II stars is (4$\pm$3)x10$^{-4}$ M$_{\odot}$ and (1$\pm$0.9)x10$^{-4}$ M$_{\odot}$, respectively, with the uncertainties indicating the median absolute deviations. Moreover, a non-negligible fraction close to 20$\%$ of group II sources fall below the rough disk mass-stellar mass correlation found for most lower-mass stars \citep[black and red lines in Fig. \ref{stellar_disk_mass}; see e.g.][and references therein]{Andrews_2013, Testi22}.

A similar result indicating smaller M$_{\rm dust}$ in group II HAeBes was recently obtained by \citet{Stapper_2021} from ALMA data. In addition to this result, \citet{Stapper_2021} put forward an alternative explanation. This is that group I and group II sources may have similar M$_{\rm dust}$ but different spatial dust grain distributions. In particular, they hypothesized that in group II sources most of the dust is concentrated in the inner, compact regions of the disks. As a consequence, such regions would be optically thick, trapping the radiation and leading to the inference of a lower M$_{\rm dust}$ than they actually have. In contrast, the inner parts of the disks around group I sources would be optically thin due to presence of a cavity, allowing the radiation to escape and yielding larger M$_{\rm dust}$ than those in group II. In turn, \citet{Stapper_2021} argued that the different dust grain distributions may be associated to the presence of giant planets in the disks of group I sources.

Therefore, regardless of the interpretation the typically stronger (sub-)mm continuum emission of group I stars can currently be understood in terms of a higher probability to host giant planets, compared to group II HAeBes.

\subsubsection{Planet candidates, metallicities and SED groups}\label{planets}

The hypothesis that giant planets are more likely present in disks around group I sources with low stellar metallicities than in the rest of the HAeBes can only be unambiguously tested once a statistically significant sample of such planets are detected. However, confirming potential detections of forming planets in protoplanetary disks is still a challenging task \citep[e.g.][]{Mendigutia_2018} and  has only been possible around the low-mass T Tauri star PDS 70 \citep{Keppler_2018,Haffert2019}. In addition, a possible forming planet has been recently discovered around the HAeBe star AB Aur \citep[][see also \citealt{Zhou2022}]{Currie_2022}. The properties of this system are fully consistent with the above hypothesis: AB Aur belongs to group I \citep{Guzman-Diaz_2021} and we derive a relatively small [M/H] = -0.35 $\pm$ -0.25 
\citep[based on the individual abundances in][and the procedure indicated in Sect. \ref{comp}]{Kama15}.

In the rest of this section we aim to provide a rough quantification about the possible presence of giant planets in relation with the SED groups and the stellar metallicities, based on observed disk structures from high-resolution techniques. It must be noted that such a quantification is difficult to assess, given that most -if not all- structures observed in protoplanetary disks can be potentially explained either invoking the presence of planets, or different phenomena \citep[for a recent example see e.g.][]{Demidova2022}.

An important exception may be the "velocity kinks", perturbations of the keplerian orbits of the gas, most probably caused by the presence of giant planets \citep[][but see also \citealt{Norfolk_2022}]{Pinte2018,Teague2018}. Using such a technique \citet{Pinte_2020} reported two HAeBes with plausible planets in their disks: HD 143006 and HD 163296. The previous sample can be complemented by the one in \citet{Asensio_2021}, where the best candidate young stars hosting planets based on SPHERE data were analyzed. In that work the masses of possible giant planets capable of generating the structures observed in scattered light (rings, cavities, spirals) were estimated for the HAeBes HD 100546, HD 135344B, HD 139614, HD 97048, HD 169142, CQ Tau, HD 34282 and HD 36112.

One of the most striking features when grouping together the HAeBes of both previous works -plus AB Aur making a total of 11 stars- is that all sources, except HD 163296, are classified as group I \citep{Guzman-Diaz_2021}. Considering the previous numbers alone, the probability of detecting disks with substructures presumably related to the presence of giant planets is $\sim$ 10 times larger in group I sources than in group II stars. However, the previous estimate must be considered as an upper limit, given that high-resolution observations are still biased towards brighter and more extended group I sources \citep{Garufi18,Garufi2022}. Concerning [M/H], none of the mentioned stars exceed the solar one except HD 163296 with [M/H] $\sim$ 0.2 (Table \ref{Table2}) and HD 36112 with [M/H] $\sim$ 0.09 \citep[as inferred from the values in][and the procedure of Sect. \ref{comp}]{Kama15}.

The crude estimate presented above suggests that current high-resolution data indeed support that group I sources with relatively small values of [M/H] are better candidates to host giant planets than the rest of the HAeBes.

 %-----------------------------------------------------------------
\section{Conclusions and final remarks}
\label{conclusions}

We have derived estimates of [M/H] for 67 HAeBe stars. This is, to our knowledge, the largest sample of such objects for which these values have been derived homogeneously, complementing our previous work on the stellar and circumstellar properties of HAeBes. Values of $\log g$ and $v \sin i$ are additional byproducts of the analysis. Our new data have served to robustly confirm the proposal presented by \citet{Kama15}, namely, that HAeBes with group I SEDs show typically lower [M/H] values than those of group II. The hypothesis put forward by \citet{Kama15} to explain this observational fact is that the presence of dust traps and cavities, potentially caused by Jovian planets accreting the refractory elements, are more frequent in disks of group I HAeBes. We provide further evidence that reinforces this hypothesis:

\begin{itemize}
    \item The [M/H] values of the HAeBes in our sample and the distinction between group I and II sources do not depend on their location within the galaxy, suggesting that the observed differences in [M/H] are not caused by the local environment but is most probably connected with the cavities associated to group I stars. In contrast, we do not find any relation between [M/H] and SED-based transitional disks without infrared excess at least up to $\sim$ 2.2 $\mu$m, indicating that not all mechanisms causing the associated holes -like photoevaporation- affect the abundances of the central stars.
    
    \item We confirm that the (sub-) mm continuum emission of group I stars is typically larger than that of group II. Regardless of the interpretation in terms of larger disk masses or different dust grain distributions, the previous is in line with the hypothesis that group I/low [M/H] sources are better candidates to host giant planets than the rest of the HAeBes 
    
    \item Current high-resolution imaging data available from the literature support the hypothesis too, with the best giant planet hosting candidates being mainly group I sources with relatively low metallicities.
\end{itemize}

It is noted that our results do not rule out the presence of giant planets in disks around group II, high metallicity sources. Strictly speaking, our results suggest that giant planets may trap the refractory material more frequently in group I sources than in group II, at least during the  evolutionary stage represented by each group. Nevertheless, exoplanet synthesis models indicate that whereas the effect of metallicity in forming planets is relevant for low-mass stars, giant planet formation can occur in a low metallicity (low dust-to-gas ratio) but high-mass protoplanetary disks surrounding higher-mass sources \citep{Alibert2011}. Indeed, \cite{Maldonado2016} reported that giant stars with planets do not show the metal-rich signature. This fact could be explained by the more massive protoplanetary disks of their progenitors, since these sources are more massive than the typical FGK stars that are the main focus of the planet searches around main sequence stars.

An unambiguous confirmation of the relation between group I/low [M/H] HAeBes and the presence of giant planets in the process of formation would represent a major step towards our understanding about planet frequencies around A an B stars, bridging the gap with the FGK stars mainly explored by exoplanet surveys. The current situation is ideal for that purpose, with hundreds of HAeBes recently identified and thousands of new potential members thanks to Gaia data. However, such a confirmation requires the actual detection of forming planets around HAeBes, currently limited to the candidate around the group I/ low [M/H] star AB Aur.

\begin{acknowledgements}
JG-D and IM are funded by a RyC2019-026992-I grant. JG-D, BM, IM and EV acknowledge
support from the "On the rocks II project" under grant PGC2018-101950-B-I00, and
MDM-2017-0737 Unidad de Excelencia "Maria de Maeztu"-Centro de Astrobiología (INTA-CSIC),
both funded by the Spanish Ministry of Science and Innovation/State Agency of Research
MCIN/AEI. This research is based on data obtained from the ESO Science Archive Facility
with DOIs: https://doi.org/10.18727/archive/27, https://doi.org/10.18727/archive/33,
https://doi.org/10.18727/archive/50, https://doi.org/10.18727/archive/71. The authors
acknowledge Carlos Eiroa for providing insightful suggestions based on a preliminary
version of the manuscript. The authors also acknowledge the referee for her/his
useful comments, which have served to improve the original manuscript.
\end{acknowledgements}

\bibliographystyle{aa}
\bibliography{biblio.bib}

\begin{appendix}

\onecolumn

\section{Tables}\label{appA}

\begin{table}[h]
\caption{Sample and properties}
\label{Table1}
\renewcommand{\arraystretch}{1.4}
\center
\resizebox{16.6 cm}{!} {
\begin{tabular}{l c c c c c c c c}
\hline
\hline
  \multicolumn{1}{c}{Object} &
  \multicolumn{1}{c}{RA} &
  \multicolumn{1}{c}{DEC} &
  \multicolumn{1}{c}{T{\tiny eff}} &
  \multicolumn{1}{c}{Log(L$_{*}$)} &
  \multicolumn{1}{c}{M$_{*}$} &
  \multicolumn{1}{c}{Age} &
  \multicolumn{1}{c}{Meeus group} &
  \multicolumn{1}{c}{JHK group} \\
  \multicolumn{1}{c}{} &
  \multicolumn{1}{c}{(h:m:s)} &
  \multicolumn{1}{c}{(d:m:s)} &
  \multicolumn{1}{c}{(K)} &
  \multicolumn{1}{c}{(L$_{\odot}$)} &
  \multicolumn{1}{c}{(M$_{\odot}$)} &
  \multicolumn{1}{c}{(Myr)} &
  \multicolumn{1}{c}{} &
  \multicolumn{1}{c}{} \\
\hline
  PDS 2 & 01 17 43.5 & -52 33 31 & 6750$\pm$125 & 0.77$_{-0.01}^{+0.01 }$ & 1.46$_{-0.01}^{+0.04}$ & 15.04$_{-2.50}^{+2.67}$ & I & J\\
  HD 9672 & 01 34 37.9 & -15 40 35 & 9000$\pm$125 & 1.20$_{-0.02}^{+0.02}$ & 1.93$_{-0.03}^{+0.02}$ & 11.02$_{-1.15}^{+8.96}$ & II & > Ks\\
  HD 31648 & 04 58 46.3 & +29 50 37 & 8000$\pm$125 & 1.22$_{-0.01}^{+0.01}$ & 1.85$_{-0.01}^{+0.04}$ & 7.71$_{-0.38}^{+0.23}$ & II & J\\
  UX Ori & 05 04 30.0 & -03 47 14 & \textit{8500$\pm$250} & 1.12$_{-0.20}^{+0.14}$ & 1.91$_{-0.00}^{+0.04}$ & 9.84$_{-0.00}^{+0.17}$ & II & J\\
  HD 34282 & 05 16 00.5 & -09 48 35 & \textit{9500$\pm$250} & 1.16$_{-0.02}^{+0.02}$ & < 1.90 & < 19.87 & I & J\\
  HD 290380 & 05 23 31.0 & -01 04 24 & 6250$\pm$125 & 0.81$_{-0.01}^{+0.01}$ & 1.59$_{-0.06}^{+0.06}$ & 9.32$_{-1.31}^{+0.75}$ & II & J\\
  HD 287823 & 05 24 08.0 & +02 27 47 & \textit{8375$\pm$125} & 1.08$_{-0.01}^{+0.01}$ & 1.83$_{-0.03}^{+0.04}$ & 10.56$_{-0.62}^{+3.36}$ & I & J\\
  V346 Ori & 05 24 42.8 & +01 43 48 & \textit{7750$\pm$250} & 0.86$_{-0.01}^{+0.01}$ & 1.65$_{-0.04}^{+0.04}$ & 16.21$_{-5.51}^{+3.37}$ & I & H\\
  CO Ori & 05 27 38.3 & +11 25 39 & 6500$\pm$215 & 1.36$_{-0.17}^{+0.12}$ & 2.30$_{-0.35}^{+0.30}$ & 3.92$_{-1.19}^{+2.07}$ & II & J\\
  HD 35929 & 05 27 42.8 & -08 19 39 & \textit{7000$\pm$250} & 1.97$_{-0.02}^{+0.02}$ & 3.53$_{-0.13}^{+0.08}$ & 1.20$_{-0.16}^{+0.30}$ & II & Ks\\
  HD 290500 & 05 29 48.1 & -00 23 43 & \textit{9500$\pm$500} & 1.10$_{-0.07}^{+0.06}$ & 1.85$_{-0.00}^{+0.05}$ & < 19.92 & I & H\\
  HD 244314 & 05 30 19.0 & +11 20 20 & \textit{8500$\pm$250} & 1.29$_{-0.02}^{+0.02}$ & 2.12$_{-0.07}^{+0.04}$ & 6.99$_{-0.00}^{+0.63}$ & II & J\\
  HD 244604 & 05 31 57.3 & +11 17 41 & \textit{9000$\pm$250} & 1.53$_{-0.02}^{+0.02}$ & 2.16$_{-0.01}^{+0.04}$ & 5.08$_{-0.08}^{+0.16}$ & II & J\\
  RY Ori & 05 32 09.9 & -02 49 47 & 6250$\pm$194 & 0.80$_{-0.11}^{+0.09}$ & 1.58$_{-0.17}^{+0.17}$ & 9.42$_{-2.44}^{+2.19}$ & II & J\\
  HD 36917 & 05 34 47.0 & -05 34 15 & 11500$\pm$144 & 2.61$_{-0.03}^{+0.03}$ & 4.36$_{-0.16}^{+0.08}$ & 0.89$_{-0.09}^{+0.09}$ & II & J\\
  HD 245185 & 05 35 09.6 & +10 01 51 & \textit{10000$\pm$500} & 1.48$_{-0.02}^{+0.02}$ & 2.24$_{-0.00}^{+0.05}$ & 7.08$_{-0.00}^{+0.06}$ & I & J\\
  NV Ori & 05 35 31.4 & -05 33 09 & 7000$\pm$125 & 1.33$_{-0.03}^{+0.03}$ & 2.09$_{-0.09}^{+0.06}$ & 5.04$_{-0.25}^{+0.85}$ & I & H\\
  T Ori & 05 35 50.5 & -05 28 35 & \textit{9000$\pm$500} & 1.77$_{-0.12}^{+0.09}$ & 2.52$_{-0.21}^{+0.19}$ & 3.82$_{-0.82}^{+0.59}$ &  & J\\
  CQ Tau & 05 35 58.5 & +24 44 54 & 6750$\pm$125 & 0.82$_{-0.01}^{+0.01}$ & 1.50$_{-0.01}^{+0.01}$ & 12.42$_{-1.43}^{+2.40}$ & I & H\\
  HD 37258 & 05 36 59.3 & -06 09 16 & \textit{9750$\pm$500} & 1.41$_{-0.02}^{+0.02}$ & 2.27$_{-0.08}^{+0.03}$ & 5.94$_{-0.78}^{+1.01}$ & II & J\\
  HD 290770 & 05 37 02.4 & -01 37 21 & \textit{10500$\pm$250} & 1.74$_{-0.01}^{+0.01}$ & 2.64$_{-0.04}^{+0.13}$ & 4.01$_{-0.04}^{+0.39}$ & II & J\\
  BF Ori & 05 37 13.3 & -06 35 01 & \textit{9000$\pm$250} & 1.13$_{-0.12}^{+0.09}$ & 1.85$_{-0.00}^{+0.15}$ & 17.14$_{-0.00}^{+2.80}$ & II & J\\
  HD 37357 & 05 37 47.1 & -06 42 30 & \textit{9500$\pm$250} & 1.94$_{-0.10}^{+0.08}$ & 2.80$_{-0.20}^{+0.20}$ & 2.96$_{-0.74}^{+0.44}$ & II & J\\
  HD 290764 & 05 38 05.3 & -01 15 22 & \textit{7875$\pm$375} & 1.34$_{-0.02}^{+0.02}$ & 1.99$_{-0.04}^{+0.03}$ & 6.10$_{-0.12}^{+0.62}$ & I & J\\
  V599 Ori & 05 38 58.6 & -07 16 46 & \textit{8000$\pm$250} & 1.49$_{-0.07}^{+0.06}$ & 2.15$_{-0.10}^{+0.11}$ & 5.09$_{-0.51}^{+0.91}$ & I & J\\
  V350 Ori & 05 40 11.8 & -09 42 11 & \textit{9000$\pm$250} & 0.96$_{-0.43}^{+0.21}$ & < 1.91 & < 15.03 & II & J\\
  HD 38120 & 05 43 11.9 & -04 59 50 & 11500$\pm$125 & 1.85$_{-0.02}^{+0.02}$ & 2.80$_{-0.04}^{+0.04}$ & 3.95$_{-0.58}^{+0.08}$ & I & J\\
  HD 39014 & 05 44 46.3 & -65 44 08 & 8000$\pm$125 & 1.66$_{-0.01}^{+0.01}$ & 2.48$_{-0.06}^{+0.07}$ & 3.52$_{-0.31}^{+0.28}$ & II & > Ks\\
  PDS 124 & 06 06 58.5 & -05 55 07 & \textit{10250$\pm$250} & 1.59$_{-0.02}^{+0.02}$ & 2.38$_{-0.05}^{+0.02}$ & 5.96$_{-0.75}^{+0.14}$ & I & H\\
  LkHa 339 & 06 10 57.8 & -06 14 40 & \textit{10500$\pm$250} & 1.62$_{-0.01}^{+0.01}$ & 2.52$_{-0.11}^{+0.08}$ & 4.91$_{-0.31}^{+0.21}$ & I & H\\
  HBC 217 & 06 40 42.2 & +09 33 37 & 6000$\pm$125 & 0.82$_{-0.01}^{+0.01}$ & 1.75$_{-0.10}^{+0.10}$ & 6.89$_{-1.55}^{+1.13}$ & I & J\\
  HBC 222 & 06 40 51.2 & +09 44 46 & 6500$\pm$174 & 0.97$_{-0.07}^{+0.06}$ & 1.70$_{-0.11}^{+0.12}$ & 8.24$_{-0.00}^{+1.77}$ &  & J\\
  PDS 130 & 06 49 58.6 & -07 38 52 & \textit{10500$\pm$250} & 1.90$_{-0.03}^{+0.03}$ & 2.74$_{-0.13}^{+0.10}$ & 3.08$_{-0.08}^{+0.22}$ & I & J\\
  HD 68695 & 08 11 44.6 & -44 05 09 & \textit{9250$\pm$250} & 1.35$_{-0.02}^{+0.02}$ & 2.08$_{-0.03}^{+0.06}$ & 8.02$_{-0.74}^{+0.91}$ & I & H\\
  GSC 8581-2002 & 08 44 23.6 & -59 56 58 & \textit{9750$\pm$250} & 1.52$_{-0.01}^{+0.01}$ & 2.40$_{-0.09}^{+0.00}$ & 5.12$_{-0.12}^{+0.86}$ & I & > Ks\\
  PDS 33 & 08 48 45.7 & -40 48 21 & \textit{9750$\pm$250} & 1.33$_{-0.01}^{+0.01}$ & < 2.10 & < 9.19 & I & J\\
  PDS 297 & 09 42 40.3 & -56 15 34 & \textit{10750$\pm$250} & 2.25$_{-0.02}^{+0.02}$ & 3.36$_{-0.10}^{+0.04}$ & 1.97$_{-0.05}^{+0.02}$ &  & > Ks\\
  HD 87403 & 10 02 51.4 & -59 16 55 & \textit{10000$\pm$250} & 2.83$_{-0.04}^{+0.04}$ & 5.54$_{-0.26}^{+0.23}$ & 0.40$_{-0.06}^{+0.05}$ &  & > Ks\\
  HD 95881 & 11 01 57.6 & -71 30 48 & \textit{10000$\pm$250} & 2.97$_{-0.02}^{+0.02}$ & 6.40$_{-0.19}^{+0.07}$ & 0.23$_{-0.03}^{+0.04}$ & II & J\\
  HD 97048 & 11 08 03.2 & -77 39 17 & \textit{10500$\pm$500} & 1.81$_{-0.01}^{+0.01}$ & 2.80$_{-0.03}^{+0.03}$ & 3.90$_{-0.60}^{+0.07}$ & I & J\\
  HD 98922 & 11 22 31.7 & -53 22 11 & \textit{10500$\pm$250} & 3.16$_{-0.02}^{+0.02}$ & 7.01$_{-0.02}^{+0.11}$ & 0.20$_{-0.01}^{+0.01}$ & II & J\\
  HD 100453 & 11 33 05.5 & -54 19 29 & \textit{7250$\pm$250} & 0.79$_{-0.01}^{+0.01}$ & 1.60$_{-0.04}^{+0.05}$ & 19.28$_{-0.68}^{+0.70}$ & I & J\\
\hline
\end{tabular}
}
\end{table}

\begin{table}[h]
\renewcommand{\arraystretch}{1.4}
\center
\resizebox{16.1cm}{!} {
\begin{tabular}{l c c c c c c c c}
\hline
\hline
  \multicolumn{1}{c}{Object} &
  \multicolumn{1}{c}{RA} &
  \multicolumn{1}{c}{DEC} &
  \multicolumn{1}{c}{T{\tiny eff}} &
  \multicolumn{1}{c}{Log(L$_{*}$)} &
  \multicolumn{1}{c}{M$_{*}$} &
  \multicolumn{1}{c}{Age} &
  \multicolumn{1}{c}{Meeus group} &
  \multicolumn{1}{c}{JHK group} \\
  \multicolumn{1}{c}{} &
  \multicolumn{1}{c}{(h:m:s)} &
  \multicolumn{1}{c}{(d:m:s)} &
  \multicolumn{1}{c}{(K)} &
  \multicolumn{1}{c}{(L$_{\odot}$)} &
  \multicolumn{1}{c}{(M$_{\odot}$)} &
  \multicolumn{1}{c}{(Myr)} &
  \multicolumn{1}{c}{} &
  \multicolumn{1}{c}{} \\
\hline
  HD 100546 & 11 33 25.3 & -70 11 41 & \textit{9750$\pm$500} & 1.34$_{-0.01}^{+0.01}$ & 2.10$_{-0.03}^{+0.05}$ & 7.67$_{-0.67}^{+0.36}$ & I & J\\
  HD 101412 & 11 39 44.4 & -60 10 28 & \textit{9750$\pm$250} & 1.69$_{-0.01}^{+0.01}$ & 2.39$_{-0.02}^{+0.01}$ & 4.06$_{-0.01}^{+0.08}$ & II & J\\
  HD 104237 & 12 00 04.9 & -78 11 35 & \textit{8000$\pm$250} & 1.29$_{-0.04}^{+0.04}$ & 1.90$_{-0.05}^{+0.07}$ & 6.99$_{-0.87}^{+0.43}$ & II & J\\
  HD 132947 & 15 04 56.0 & -63 07 53 & \textit{10250$\pm$250} & 1.80$_{-0.01}^{+0.01}$ & 2.77$_{-0.01}^{+0.05}$ & 3.75$_{-0.81}^{+0.04}$ & II & > Ks\\
  HD 135344B & 15 15 48.4 & -37 09 16 & \textit{6375$\pm$125} & 0.71$_{-0.01}^{+0.01}$ & 1.46$_{-0.04}^{+0.07}$ & 10.48$_{-0.49}^{+0.97}$ & I & J\\
  HD 139614 & 15 40 46.4 & -42 29 54 & \textit{7750$\pm$250} & 0.83$_{-0.01}^{+0.01}$ & 1.60$_{-0.00}^{+0.02}$ & 19.35$_{-0.00}^{+0.64}$ & I & H\\
  HD 141569 & 15 49 57.7 & -03 55 17 & \textit{9500$\pm$250} & 1.40$_{-0.01}^{+0.01}$ & 2.12$_{-0.01}^{+0.04}$ & 7.97$_{-0.03}^{+0.03}$ & II & > Ks\\
  HD 142666 & 15 56 40.0 & -22 01 40 & \textit{7250$\pm$250} & 1.13$_{-0.01}^{+0.01}$ & 1.75$_{-0.00}^{+0.03}$ & 8.73$_{-0.74}^{+0.08}$ & II & H\\
  HD 142527 & 15 56 41.9 & -42 19 24 & \textit{6500$\pm$250} & 1.35$_{-0.01}^{+0.01}$ & 2.20$_{-0.05}^{+0.05}$ & 4.40$_{-0.38}^{+0.49}$ & I & J\\
  HD 143006 & 15 58 36.9 & -22 57 16 & 5500$\pm$125 & 0.54$_{-0.02}^{+0.02}$ & 1.74$_{-0.13}^{+0.10}$ & 5.10$_{-0.00}^{+1.89}$ & I & J\\
  HD 144432 & 16 06 57.9 & -27 43 10 & \textit{7500$\pm$250} & 1.21$_{-0.01}^{+0.01}$ & 1.82$_{-0.01}^{+0.03}$ & 7.98$_{-0.21}^{+0.02}$ & II & J\\
  HR 5999 & 16 08 34.3 & -39 06 19 & \textit{8500$\pm$250} & 1.97$_{-0.09}^{+0.08}$ & 3.19$_{-0.29}^{+0.21}$ & 1.98$_{-0.41}^{+0.33}$ & II & H\\
  V718 Sco & 16 13 11.6 & -22 29 07 & \textit{7750$\pm$250} & 1.08$_{-0.04}^{+0.03}$ & 1.71$_{-0.02}^{+0.04}$ & 9.02$_{-0.11}^{+0.56}$ & II & Ks\\
  HD 149914 & 16 38 28.6 & -18 13 14 & 9500$\pm$125 & 2.05$_{-0.01}^{+0.01}$ & 3.07$_{-0.05}^{+0.07}$ & 2.02$_{-0.02}^{+0.05}$ & II & > Ks\\
  HD 150193 & 16 40 17.9 & -23 53 45 & \textit{9250$\pm$250} & 1.36$_{-0.01}^{+0.01}$ & 2.25$_{-0.08}^{+0.00}$ & 6.00$_{-0.02}^{+1.00}$ & II & J\\
  HD 158643 & 17 31 25.0 & -23 57 46 & 9500$\pm$150 & 2.25$_{-0.01}^{+0.01}$ & 3.60$_{-0.02}^{+0.07}$ & 1.35$_{-0.14}^{+0.16}$ & II & Ks\\
  HD 163296 & 17 56 21.3 & -21 57 22 & \textit{9000$\pm$250} & 1.19$_{-0.05}^{+0.04}$ & 1.91$_{-0.00}^{+0.12}$ & 10.00$_{-2.00}^{+9.50}$ & II & J\\
  HD 169142 & 18 24 29.8 & -29 46 50 & 7250$\pm$125 & 0.76$_{-0.01}^{+0.01}$ & 1.55$_{-0.00}^{+0.03}$ & < 20.00 & I & J\\
  HD 176386 & 19 01 38.9 & -36 53 27 & 9750$\pm$125 & 1.64$_{-0.01}^{+0.01}$ & 2.63$_{-0.25}^{+0.00}$ & 3.88$_{-0.01}^{+0.57}$ &  & > Ks\\
  HD 179218 & 19 11 11.3 & +15 47 15 & \textit{9500$\pm$250} & 2.02$_{-0.01}^{+0.01}$ & 2.99$_{-0.04}^{+0.01}$ & 2.35$_{-0.16}^{+0.19}$ & I & H\\
  WW Vul & 19 25 58.8 & +21 12 31 & 8500$\pm$125 & 1.41$_{-0.06}^{+0.05}$ & 2.04$_{-0.05}^{+0.06}$ & 6.02$_{-0.20}^{+0.52}$ & II & J\\
  PX Vul & 19 26 40.3 & +23 53 51 & 6500$\pm$125 & 1.49$_{-0.03}^{+0.03}$ & 2.59$_{-0.18}^{+0.13}$ & 2.95$_{-0.64}^{+0.20}$ & II & J\\
  V1295 Aql & 20 03 02.5 & +05 44 17 & \textit{9750$\pm$250} & 2.88$_{-0.03}^{+0.03}$ & 6.00$_{-0.21}^{+0.20}$ & 0.30$_{-0.02}^{+0.02}$ & II & J\\
  HD 199603 & 20 58 41.8 & -14 29 00 & 7500$\pm$125 & 1.39$_{-0.01}^{+0.01}$ & 2.10$_{-0.05}^{+0.04}$ & 5.19$_{-0.19}^{+0.54}$ & II & > Ks\\
  BP Psc & 23 22 24.7 & -02 13 42 & 5250$\pm$125 & 0.40$_{-0.28}^{+0.17}$ & 1.65$_{-0.43}^{+0.24}$ & 4.89$_{-2.75}^{+8.26}$ & I & J\\
\hline\end{tabular}
}
\tablefoot{Columns 1, 2, and 3 show the names of the stars and their coordinates. Column 4 lists the effective temperatures, taken from \citet{Wichittanakom20} when displayed in italics or from \citet{Guzman-Diaz_2021} otherwise. The values in the rest of the columns are taken from the latter work, listing the stellar luminosities, masses, ages, and the Meeus and JHK SED groups (see text).}
\end{table}

\begin{table}[]
\caption{Spectra used and stellar parameters derived in this work}
\label{Table2}
\renewcommand{\arraystretch}{1.4}
\center
\resizebox{12.5 cm}{!} {
\begin{tabular}{l c c c c c}
\hline
\hline
  \multicolumn{1}{c}{Object} &
  \multicolumn{1}{c}{Instruments} &
  \multicolumn{1}{c}{Resolutions} &
  \multicolumn{1}{c}{$v \sin i$} &
  \multicolumn{1}{c}{$\log g$} &
  \multicolumn{1}{c}{[M/H]}  \\
  \multicolumn{1}{c}{} &
  \multicolumn{1}{c}{I$_{1}$/I$_{2}$} &
  \multicolumn{1}{c}{R$_{1}$/R$_{2}$} &
  \multicolumn{1}{c}{(km/s)} &
  \multicolumn{1}{c}{(dex)} &
  \multicolumn{1}{c}{(dex)} \\
\hline  
  PDS 2 & H/XS & 80000/9900 & 15$\pm${1} & 4.23$\pm${0.05} & $\leq$ -0.10\\
  HD 9672 & H/H & 115000/115000 & 200$\pm${10} & 3.97$\pm${0.07} & 0.20$\pm${0.10}\\
  HD 31648 & U/XS & 71000/9900 & 102$\pm${5} & 4.13$\pm${0.05} & -0.25$\pm${0.13}\\
  UX Ori & XS/XS & 9900/9900 & 225$\pm${11} & 3.75$\pm${0.12} & 0.00$\pm${0.10}\\
  HD 34282 & XS/XS & 9900/9900 & 105$\pm${5} & 4.43$\pm${0.10} & $\leq$ 0.00\\
  HD 290380 & U/U & 71000/87400 & 75$\pm${4} & 3.98$\pm${0.05} & 0.00$\pm${0.10}\\
  HD 287823 & XS/XS & 9900/9900 & 27$\pm${1} & 4.25$\pm${0.11} & -0.50$\pm${0.14}\\
  V346 Ori & U/XS & 41000/9900 & 115$\pm${6} & 4.50$\pm${0.05} & 0.00$\pm${0.13}\\
  CO Ori & XS/XS & 9900/9900 & 55$\pm${3} & 3.76$\pm${0.05} & 0.15$\pm$0.10\\
  HD 35929 & H/XS & 80000/9900 & 60$\pm${3} & 3.50$\pm${0.05} & 0.12$\pm${0.10}\\
  HD 290500 & XS/XS & 9900/9900 & 80$\pm${4} & 3.79$\pm${0.25} & 0.00$\pm${0.10}\\
  HD 244314 & XS/XS & 9900/9900 & 55$\pm${3} & 4.08$\pm${0.05} & 0.00$\pm${0.10}\\
  HD 244604 & XS/XS & 9900/9900 & 100$\pm${5} & 3.95$\pm${0.18} & 0.50$\pm${0.10}\\
  RY Ori & XS/XS & 3300/3300 & 57$\pm${3} & 4.00$\pm${0.05} & 0.00$\pm${0.14}\\
  HD 36917 & XS/XS & 9900/9900 & 110$\pm${6} & 4.29$\pm${0.07} & 0.20$\pm$0.10\\
  HD 245185 & XS/XS & 9900/9900 & 115$\pm${6} & 4.22$\pm${0.06} & -0.50$\pm${0.30}\\
  NV Ori & XS/XS & 3300/3300 & 75$\pm${4} & 3.77$\pm${0.05} & 0.14$\pm$0.10\\
  T Ori & U/XS & 41000/9900 & 150$\pm${8} & 3.65$\pm${0.17} & 0.10$\pm${0.10}\\
  CQ Tau & XS/XS & 3300/3300 & 98$\pm$5 & 4.13$\pm${0.05} & $\leq$ 0.00\\
  HD 37258 & U/XS & 41000/9900 & 210$\pm${11} & 4.21$\pm${0.15} & 0.35$\pm${0.10}\\
  HD 290770 & XS/XS & 9900/9900 & 230$\pm${12} & 4.11$\pm${0.03} & -1.00$\pm${0.33}\\
  BF Ori & -/XS & -/9900 & 37$\pm${2} (1) & 3.87$\pm${0.34} & 0.20$\pm${0.13}\\
  HD 37357 & U/XS & 41000/9900 & 140$\pm${7} & 4.19$\pm${0.10} & 0.50$\pm${0.20}\\
  HD 290764 & XS/XS & 9900/9900 & 55$\pm${3} & 3.94$\pm${0.05} & -0.15$\pm${0.10}\\
  V599 Ori & XS/XS & 9900/9900 & 47$\pm${2} & 3.87$\pm${0.05} & -0.35$\pm${0.10}\\
  V350 Ori & XS/XS & 9900/9900 & 125$\pm${6} & 4.12$\pm${0.17} & 0.50$\pm${0.10}\\
  HD 38120 & XS/XS & 3300/3300 & 105$\pm${5} & 4.50$\pm${0.11} & $\leq$ 0.00\\
  HD 39014 & H/H & 115000/115000 & 195$\pm${10} & 3.80$\pm${0.05} & 0.20$\pm${0.20}\\
  PDS 124 & XS/XS & 9900/9900 & 140$\pm${7} & 4.26$\pm${0.13} & 0.10$\pm$0.15\\
  LkHa 339 & XS/XS & 9900/9900 & 130$\pm${7} & 4.15$\pm${0.14} & 0.50$\pm$0.12\\
  HBC 217 & U/U & 48000/48000 & 36$\pm${1} & 3.95$\pm${0.05} & -0.25$\pm${0.10}\\
  HBC 222 & G/G & 24000/24000 & 52$\pm${5} & 3.94$\pm${0.05} & 0.00$\pm${0.10}\\
  PDS 130 & XS/XS & 9900/9900 & 100$\pm${5} & 3.77$\pm${0.20} & $\geq$ 0.20\\
  HD 68695 & XS/XS & 9900/9900 & 45$\pm${2} & 4.35$\pm${0.08} & -0.50$\pm${0.10}\\
  GSC 8581-2002 & XS/XS & 9900/9900 & 155$\pm${8} & 3.90$\pm${0.05} & 0.00$\pm${0.10}\\
  PDS 33 & XS/XS & 9900/9900 & 140$\pm${7} & 4.41$\pm${0.08} & -0.50$\pm${0.17}\\
  PDS 297 & XS/XS & 9900/9900 & 200$\pm${10} & 3.92$\pm${0.06} & 0.20$\pm$0.10\\
  HD 87403 & XS/XS & 9900/9900 & 98$\pm${5} & 3.35$\pm${0.11} & 0.00$\pm${0.10}\\
  HD 95881 & U/XS & 41000/9900 & 70$\pm${4} & 3.19$\pm${0.11} & 0.00$\pm${0.10}\\
  HD 97048 & U/XS & 71000/9900 & 145$\pm${7} & 4.28$\pm${0.10} & -0.50$\pm${0.13}\\
  HD 98922 & U/XS & 71000/9900 & 43$\pm${2} & 3.61$\pm${0.05} & 0.00$\pm${0.13}\\
  HD 100453 & H/XS & 115000/9900 & 50$\pm$3 & 4.47$\pm${0.05} & -0.10$\pm$0.10\\
  
\hline
\end{tabular}
}
\end{table}

\begin{table}[h]
\renewcommand{\arraystretch}{1.4}
\center
\resizebox{12 cm}{!} {
\begin{tabular}{l c c c c c}
\hline
\hline
  \multicolumn{1}{c}{Object} &
  \multicolumn{1}{c}{Instruments} &
  \multicolumn{1}{c}{Resolutions} &
  \multicolumn{1}{c}{$v \sin i$} &
  \multicolumn{1}{c}{$\log g$} &
  \multicolumn{1}{c}{[M/H]} \\
  \multicolumn{1}{c}{} &
  \multicolumn{1}{c}{I$_{1}$/I$_{2}$} &
  \multicolumn{1}{c}{R$_{1}$/R$_{2}$} &
  \multicolumn{1}{c}{(km/s)} &
  \multicolumn{1}{c}{(dex)} &
  \multicolumn{1}{c}{(dex)} \\
\hline  
\hline
  HD 100546 & U/XS & 71000/9900 & 58$\pm${3} & 4.01$\pm${0.03} & $\leq$ -1.00\\
  HD 101412 & U/U & 71000/107200 & 3$\pm$1 & 4.32$\pm${0.19} & $\leq$ 0.00\\
  HD 104237 & U/U & 65000/74500 & 12$\pm$1 & 3.88$\pm${0.05} & $\leq$ 0.20\\
  HD 132947 & U/XS & 41000/9900 & 118$\pm${6} & 3.95$\pm${0.11} & 0.20$\pm${0.10}\\
  HD 135344B & U/XS & 71000/9900 & 70$\pm${4} & 4.15$\pm${0.05} & 0.00$\pm${0.13}\\
  HD 139614 & XS/XS & 9900/9900 & 32$\pm${2} & 4.45$\pm${0.05} & -0.25$\pm${0.10}\\
  HD 141569 & U/XS & 71000/9900 & 220$\pm${11} & 4.09$\pm${0.08} & -0.50$\pm${0.38}\\
  HD 142666 & U/XS & 41000/9900 & 65$\pm$3 & 3.99$\pm${0.05} & 0.00$\pm${0.10}\\
  HD 142527 & U/XS & 41000/9900 & 48$\pm$2 & 3.72$\pm${0.05} & 0.08$\pm$0.13\\
  HD 143006 & XS/XS & 9900/9900 & 15$\pm$1 & 4.10$\pm${0.05} & $\leq$ -0.15\\
  HD 144432 & U/XS & 41000/9900 & 77$\pm${4} & 4.00$\pm${0.05} & 0.15$\pm$0.10\\
  HR 5999 & U/XS & 41000/9900 & 180$\pm${9} & 3.50$\pm${0.11} & 0.20$\pm${0.10}\\
  V718 Sco & H/XS & 115000/5500 & 120$\pm${6} & 4.27$\pm${0.05} & 0.00$\pm${0.10}\\
  HD 149914 & XS/XS & 3300/3300 & 200$\pm${10} & 3.50$\pm${0.11} & 0.00$\pm${0.14}\\
  HD 150193 & U/XS & 41000/9900 & 107$\pm${5} & 4.08$\pm${0.14} & 0.20$\pm${0.10}\\
  HD 158643 & XS/XS & 5500/5500 & 222$\pm${11} & 3.50$\pm${0.11} & 0.20$\pm${0.10}\\
  HD 163296 & U/XS & 71000/9900 & 126$\pm${6} & 4.06$\pm${0.20} & 0.20$\pm${0.10}\\
  HD 169142 & H/H & 115000/115000 & 48$\pm${2} & 4.34$\pm${0.05} & -0.50$\pm${0.33}\\
  HD 176386 & U/XS & 71000/9900 & 170$\pm${9} & 3.70$\pm${0.07} & 0.20$\pm${0.10}\\
  HD 179218 & XS/XS & 5500/5500 & 75$\pm${4} & 3.90$\pm${0.04} & -0.50$\pm${0.17}\\
  WW Vul & U/U & 41000/42300 & 190$\pm${10} & 3.65$\pm${0.01} & $\geq$ 0.20\\
  PX Vul & U/U & 41000/42300 & 78$\pm${4} & 3.63$\pm${0.05} & 0.00$\pm${0.10}\\
  V1295 Aql & XS/XS & 9900/9900 & 25$\pm${1} & 3.69$\pm${0.11} & $\geq$ 0.10\\
  HD 199603 & U/U & 41000/42300 & 90$\pm${5} & 3.89$\pm${0.05} & 0.00$\pm${0.10}\\
  BP Psc & H/H & 115000/115000 & 40$\pm${2} & 4.16$\pm${0.05} & 0.20$\pm${0.16}\\
  
\hline
\end{tabular}
}
\tablefoot{Column 2 indicates the instruments I$_1$ and I$_2$ used to derive the projected rotational velocities (Col. 4, see the references below for previous estimates of this parameter), and the widths of the Balmer lines and [M/H] values (related to Cols. 5 and 6), respectively. XS refers to XSHOOTER, U to UVES, H to HARPS, and G to GIRAFFE. Column 3 shows the resolutions at $\sim$ 5000 \r{A} of the spectra acquired with the aforementioned instruments.}
\tablebib{(1) \citet{Mora_2001}.}
\end{table}

\twocolumn

\section{Notes on individual objects}\label{appB}

\begin{itemize}
    
    \item \textbf{HD 31648:} Discrepancies in stellar parameters lead to different values of [M/H] in the literature. For instance, in \citet{Folsom_2012} a [M/H] = 0.23 is found for a T{\tiny eff} = 8800 K and $\log g$ = 4.1. On the other hand, in \citet{Montesinos09} they estimate a [M/H] = 0.0, being in this case T{\tiny eff} = 8250 K and $\log g$ = 4.0. In this work we derive a surface gravity similar to that of these works ($\log g$ = 4.13) but a lower effective temperature is used (T{\tiny eff} = 8000 K), deriving finally a [M/H] = -0.25.
    \item \textbf{HD 290380:} The observed spectrum analyzed has a SNR below 100 (but high enough; $\geq$ 50).
    \item \textbf{HD 287823:} A $\log g$ = 4.25, which is similar to that provided by \citealt{Wichittanakom20}, was fixed. The assumed uncertainty in this parameter was the mean error of $\log g$ from HAeBes with T{\tiny eff} $>$ 8000 K, which is 0.11.
    \item\textbf{BF Ori:} The $v \sin i$ of \citet{Mora_2001} was considerered since the value of such parameter estimated from the process described in Sect. \ref{vrot} does not fit properly the profiles of the photospheric lines of the selected regions in the interval 5000-7000 \r{A}. 
    \item\textbf{V599 Ori:} The observed spectrum analyzed has a SNR below 100 (but high enough; $\geq$ 70).
    \item \textbf{HD 38120:} The widths of the Balmer lines in the observed spectrum are larger than those estimated in the model generated with $\log g$ = 4.50. We assumed this value of $\log g$, and the same error as HD 287823 for this object.
    \item \textbf{HD 87048}: The value of $\log g$ for each [M/H] was estimated trough PARSEC V2.1s evolutionary tracks giving the T{\tiny eff} from \citet{Wichittanakom20}, and stellar mass and radius from \citet{Guzman-Diaz_2021}. The considered error in $\log g$ is the same as HD 287823.
    \item \textbf{HD 95881}:  The value of $\log g$ for each [M/H] was estimated trough PARSEC V2.1s evolutionary tracks giving the T{\tiny eff} from \citet{Wichittanakom20}, and stellar mass and radius from \citet{Guzman-Diaz_2021}. The considered error in $\log g$ is the same as HD 287823.
    \item \textbf{HD 141569:} The observed spectrum was challenging to analyze due to the scarcity of spectral lines, the high $v \sin i$, and the signal to noise. Similar $\chi^{2}$ were obtained in the fits with models generated at [M/H] equal to -0.5 and 0.0. Nevertheless, for this source we decided to fix [M/H] = -0.5 as this value is similar to the one derived from \citet{Kama15}, which is [M/H] = -0.69.
    \item \textbf{HR 5999:} The widths of the Balmer series lines in the observed spectra are lower than those obtained in the models generated at $\log g$ = 3.5. The $\log g$ for each [M/H] was also estimated from PARSEC V2.1s evolutionary tracks, but no values below 3.5 were obtained. Therefore, a $\log g$ = 3.5 with the same error as HD 287823 were adopted. 
    \item \textbf{HD 149914:} A $\log g$ = 3.5 was assumed due to the same reason described in HR 5999.
    \item \textbf{HD 158643:} A $\log g$ = 3.5 was assumed due to the same reason described in the case of HR 5999.
    \item \textbf{PX Vul:} The observed spectrum analyzed has a SNR below 100 (but high enough; $\sim$ 50).
    \item \textbf{BP Psc:} The observed spectrum analyzed has a SNR below 100 (but high enough; $\geq$ 70).
    
\end{itemize}

\end{appendix}

\end{document}